\documentclass[aps,
prb,
reprint,
groupedaddress,
amsmath,
amssymb,
showkeys,
nofootinbib]{revtex4-2}

\usepackage{graphicx}
\usepackage{dcolumn}
\usepackage{enumitem}
\usepackage{bm}
\usepackage{physics}
\usepackage{tensor}
\usepackage{slashed}
\usepackage[normalem]{ulem}
\usepackage[linktocpage=true]{hyperref}
\hypersetup{
    colorlinks,
    linkcolor={blue},
    urlcolor={blue},
    citecolor={blue},
    breaklinks=true,
}
\usepackage{color}
\numberwithin{equation}{section}
\renewcommand{\theequation}{\arabic{section}.\arabic{equation}}

\newcommand{\del}{\partial}
\newcommand{\nn}{\nonumber}

\begin{document}

\title{Vestigial Nematic Order at Zero Temperature in Two-Dimensional Frustrated Quantum Antiferromagnets}

\author{Matthew C. O'Brien}
\email{mco5@illinois.edu}
\author{Eduardo Fradkin}%
\email{efradkin@illinois.edu}
\affiliation{%
Department of Physics and Anthony J. Leggett Institute for Condensed Matter Theory, Grainger College of Engineering, University of Illinois Urbana-Champaign, 1110 West Green Street, Urbana, IL 61801, USA
}%

\date{\today}

\begin{abstract}
The phase diagram of the two-dimensional quantum $J_1$-$J_3$ Heisenberg antiferromagnet on a square lattice is a long-standing open problem. Despite recent advances in numerical techniques for quantum spin models, a detailed analytical theory is still lacking. We address this problem using a semiclassical approach based on a continuum nonlinear sigma model effective field theory, applying the nonperturbative large-$N$ technique to map out the phase diagram and determine the magnetic correlations. We show that previously-overlooked interactions are crucial for stabilizing a vestigial nematic phase, both at finite and zero temperature. Our results reveal that the spontaneous breaking of global symmetries in the $J_1$-$J_3$ model is controlled by the strength of infrared quantum fluctuations which are enhanced by proximity to the classical Lifshitz point.
\end{abstract}

\maketitle

\section{Introduction \label{sec:intro}}

The nature of quantum disordered phases of matter in magnetic systems has intrigued the condensed matter physics community for decades. Key themes that have emerged over the years are the importance of frustrating interactions in destroying long-range magnetic order and the possibility of finding exotic physics at zero temperature in the vicinity of classical phase transitions. A canonical model of frustrated antiferromagnetism which unites these two themes is the $SU(2)$-invariant two-dimensional (2D) quantum $J_1$-$J_2$-$J_3$ Heisenberg model on the square lattice, which has the Hamiltonian
\begin{equation}
    H = J_1 \sum_{\langle i,j\rangle_1} \vb{S}_i \cdot \vb{S}_j + J_2 \sum_{\langle i,j\rangle_2} \vb{S}_i \cdot \vb{S}_j + J_3 \sum_{\langle i,j\rangle_3} \vb{S}_i \cdot \vb{S}_j , \label{eq:j1j2j3}
\end{equation}
where $\vb{S}_i = (S_i^{(x)}, S_i^{(y)}, S_i^{(z)})$ is the spin-$S$ operator at site $i$, $J_1$, $J_2$, $J_3>0$ are the antiferromagnetic exchange coupling constants between pairs of nearest neighbor, second-nearest neighbor, and third-nearest neighbor sites, respectively, and $\langle i, j \rangle_k$ denotes summation over the $k$\textsuperscript{th} nearest neighbors. The phenomenological richness of this model comes from the fact that each of these interactions frustrate the others; no two pairs of interactions can be individually minimized simultaneously, even on the square lattice. Further, it has been shown rigorously from the Lieb-Schultz-Mattis theorem and its generalizations \cite{Lieb1961,Oshikawa2000,Hastings2004}, that at zero temperature, the ground state of a system with a single spin-1/2 degree of freedom in each unit cell must be either gapless (e.g., magnetically ordered) or gapped and degenerate (e.g., the topological degeneracy of a $\mathbb{Z}_2$ spin liquid). Therefore, featureless paramagnets are forbidden at zero temperature for $S=1/2$ and an exotic ground state is guaranteed.

Over the last several decades, a tremendous amount of work has been done on the special case when $J_3 = 0$, which is the $J_1$-$J_2$ model made famous by Anderson's resonating valence bond (RVB) model and its application to the cuprates \cite{Anderson1973,Anderson1987,Kivelson1987TopologySuperconductivity} (see, for example, Ref. \cite{Savary2017} and the references therein for a review of the extensive work done), and questions about the nature of its phase transitions remain active areas of research \cite{Wang2016,Poilblanc2017,Wang2018,Ferrari2020,Nomura2021,Liu2022a,Qian2024,Ruckriegel2024}. In this work, however, we are interested in the opposite limit, where $J_2 = 0$. This case has also been discussed for quite some time \cite{Ioffe1988,Leung1996,Capriotti2004a,Capriotti2004,Ralko2009,Sindzingre2010,Reuther2011,Kharkov2018,Kharkov2020,Wu2022,Liu2022,Liu2024}, though more sparsely, and the zero temperature quantum phase diagram as a function of frustration $J_3$ is still not fully understood.

Classically, the $J_1$-$J_3$ model has two phases separated by the Lifshitz point (LP) at $J_3=J_1 /4$: For $J_3 < J_1 / 4$ the system is a $(\pi,\pi)$ N\'eel antiferromagnet, while for $J_3  > J_1 /4$, the staggered magnetization spontaneously breaks lattice translation and rotation symmetries by forming a unidirectional incommensurate spiral (helimagnetic antiferromagnet) with one of two wavevectors $\vb{Q}_\pm = (\pi,\pi) - (\pm Q,Q)$, where $\cos(Q) = J_1/4J_3$ \cite{Gelfand1989,Moreo1990IncommensurateModels}. At finite temperature, the Mermin-Wagner theorem guarantees the destruction of both types of long-range magnetic order in 2D by thermal fluctuations \cite{Mermin1966}. However, it was shown using classical Monte Carlo simulations that in the vicinity of the $T=0$ spiral phase, the discrete $C_{4v}\to\mathbb{Z}_2$ spontaneous symmetry breaking associated with the spiral order persists up to a nonzero critical temperature \cite{Capriotti2004}. 

The phenomenon described in Ref. \cite{Capriotti2004} is in fact a generic feature of systems with multi-component order parameters whose ordered states break several symmetries: The primary (or parent) order can be destroyed by fluctuations while a composite (bilinear or higher) order parameter can remain ordered, breaking only a subgroup of the larger symmetry. In an early example, it was shown that the magnetic order of the $(\pi,0)$ and $(0,\pi)$ stripe ground states of the $J_1$-$J_2$ model at large $J_2$ can be melted by thermal fluctuations while retaining genuine long-range directional ordering \cite{Chandra1990a}. In the modern terminology, this is known as a \textit{vestigial order} (see, for example, Refs. \cite{Nie2014,Fernandes-2019}), and should not be confused with the concept of short-range correlations of the parent order.

In the case of the $J_1$-$J_3$ model, a ``vestige'' of the zero temperature spiral phase remains, and the finite temperature phase diagram is divided into two regions: (i) The featureless paramagnet above the N\'eel phase, and (ii) the vestigial Ising nematic phase above the spiral phase. In Ref. \cite{Capriotti2004}, it was argued, but not shown, that the Ising nematic phase persists down to zero temperature, where it separates valence bond solid (VBS) and $\mathbb{Z}_2$ spin liquid phases.

Recent advances in numerical techniques for quantum spin models have prompted renewed interest in the zero temperature phase diagram of the $J_1$-$J_3$ model \cite{Wu2022,Liu2022,Liu2024}. In particular, those works provide evidence for the following sequence of phases with increasing $J_3$: (i) N\'eel, (ii) quantum spin liquid (QSL), (iii) VBS, and finally (iv) spiral.\footnote{Note that the authors of Ref. \cite{Capriotti2004} state without proof that they expect VBS order followed by a $\mathbb{Z}_2$ QSL phase on increasing $J_3$, in contrast with the recent numerical results.} Additionally, both exact diagonalization \cite{Sindzingre2010}, and finite projected entangled pair state (PEPS) tensor network \cite{Liu2024}, studies have revealed short-range spiral correlations in the vicinity of the spiral ordering transition, as well as a mixed columnar-plaquette VBS phase that spontaneously breaks $C_{4v}$ rotation symmetry. However, despite the intensifying numerical work, there have been no recent analytical studies revisiting the nature of the quantum-disordered phases and the point-group symmetries they break.

In this work, we seek to address aspects of this deficit. To this end, we consider an effective field theory for the $J_1$-$J_3$ model written in terms of the staggered magnetization order parameter vector $\vec{n}$, which transforms under the global symmetry group $O(3)$ of the Heisenberg model. To investigate nematic ordering, we apply the nonperturbative large-$N$ technique by generalizing the global symmetry to $O(N)$. In the classical limit, we reproduce the finite temperature Ising nematic phase presented in Ref. \cite{Capriotti2004}, providing, to the best of our knowledge, the first field-theoretic description of this phase; finite temperature nematic ordering has previously been demonstrated on the lattice using Schwinger boson mean-field theory \cite{Mezio2013}, and a self-consistent large-$N$ approximation \cite{Schecter2017}. We map out the $N=\infty$ phase diagram of the classical theory as a function of the field theory parameters and discuss how domain walls in the nematic order parameter affect the mean-field phase boundary.

We then apply our method to the quantum system at zero temperature. Following the original treatment of the $J_1$-$J_3$ model by Ioffe and Larkin \cite{Ioffe1988}, we emphasize how the classical LP destroys long-range magnetic order: In the vicinity of the classical N\'eel to spiral transition, magnetic excitations soften substantially, leading to infrared quantum fluctuations whose strength is directly controlled by proximity to the LP. We show that despite those strong infrared fluctuations, as with thermal fluctuations at $T>0$, a vestigial Ising nematic phase exists adjacent to the spiral phase. Concretely, for any choice of the field theory parameters where nematic order exists at finite temperature in the classical limit, a corresponding nematic phase persists at zero temperature. We also find that, at least at $N=\infty$, the transition between spiral and nematic phases is continuous. Finally, we also calculate the staggered magnetization structure factor across the quantum disordered and nematic phases, finding good agreement with numerical works \cite{Liu2024}.

Our approach complements prior large-$N$ calculations which utilized a very different $N\to\infty$ limit based on a generalization of $SU(2)$ symmetry to the symplectic group $Sp(N)$ \cite{Read1991,Sachdev1991}. Unlike in those works, our approach reveals a clear separation between the emergence of incommensurate (spiral) correlations and the discrete rotation symmetry breaking (nematic) transition. This separation demonstrates that short-range spiral correlations and nematicity are related but not synonymous; the crossover to incommensurate fluctuations generically occurring before nematic symmetry breaking.

Finally, we raise some points of clarification. The precise nature of quantum disordered ground states is generally expected to depend strongly on the spin of the local moments. For example, in the 2D Affleck-Kennedy-Lieb-Tasaki (AKLT) model \cite{Affleck1987RigorousAntiferromagnets,Affleck1988}, it is known that the edge hosts gapless excitations when the spins in the bulk have $S=2$, while it is gapped for $S=4$ \cite{Katsura2010EntanglementGraphs,Lou2011EntanglementTheory,Wierschem2016DetectionCorrelator}. However, because we use a semiclassical continuum effective field theory, our method is insensitive to the microscopic spin of the local moments and cannot capture lattice-scale physics such as the discrete translation symmetry breaking in a VBS. Formally it is exact in the limit $S\to\infty$, although semiclassical intuition remains useful in most cases (the 2D AKLT model being a notable exception) even for small integer spins such as $S=1$ \cite{Haldane1983a}. When we refer to the \textit{quantum disordered} phase of the theory we mean the absence of long-range magnetic order generally, not any specific phase (for example, a QSL, a VBS, or a trivial paramagnet). Similarly, we cannot determine whether the nematic phase for the $S=1/2$ model will be the mixed columnar–plaquette VBS proposed in Ref. \cite{Liu2024}. Nevertheless, our calculations show that infrared-scale physics strongly shapes the global phase diagram and the response to experimental probes, and that both are qualitatively consistent with the results reported in that recent numerical study.

The rest of this paper is organized as follows: In Sec. \ref{sec:model} we introduce the continuum effective field theory this work is based on. We summarize its known basic properties from both the more intuitive perturbative renormalization group (RG) perspective of Ioffe and Larkin \cite{Ioffe1988}, and the nonperturbative large-$N$ technique we use throughout. In Sec. \ref{sec:classical} we study the well-understood classical finite temperature limit of the field theory to develop and test our technique. We verify the existence of the classical nematic phase and discuss the nature of the phase transition. In Sec. \ref{sec:quantumvestigial} we demonstrate the existence of the quantum vestigial nematic phase and obtain the phase diagram as well as the static staggered magnetization structure factor across three quantum disordered regimes. We also verify the nature of the quantum disordered to nematic phase transition and map the evolution of the spiral ordering transition through an exact calculation of the full effective potential. Finally, Sec. \ref{sec:disc} presents our conclusions. We also present details of additional calculations in three appendixes: Appendix \ref{app:spinwave} describes the linear spin-wave mapping between lattice and field theory parameters. Appendix \ref{app:RG} describes our perturbative RG calculation in more detail. Appendix \ref{app:saddles} provides the full expressions for the large-$N$ saddle-point equations we calculate.

\section{The \texorpdfstring{$J_1$-$J_3$}{J1-J3} Model in the Continuum \label{sec:model}}

The features of the $J_1$-$J_2$-$J_3$ model discussed above are captured by a low-energy continuum effective field theory as long as $J_2$ or $J_3$ are not large enough to induce commensurate ordering on the ultraviolet lattice scale (e.g., columnar order). This field theory can be derived from Eq. \eqref{eq:j1j2j3} in a semiclassical $1/S$ expansion \cite{Ioffe1988}, yielding the quantum nonlinear sigma model (NLSM) Lagrangian (in real time $t$ and two spatial dimensions),
\begin{align}
    \mathcal{L} = \frac{\chi_\perp}{2} (\del_t \Vec{n})^2 - \frac{\rho}{2} (\grad \Vec{n})^2 - \frac{b_1}{2}  \left[(\partial_x^2 \Vec{n})^2 + (\partial_y^2 \Vec{n})^2\right] \label{eq:lagrangian} \\
    - b_2(\partial_x^2 \Vec{n})\cdot(\partial_y^2 \Vec{n}) +\dots, \nonumber
\end{align}
where $\Vec{n} = (n_1,n_2,n_3)$ is a unit vector $\Vec{n}^2 = 1$ representing the local staggered magnetization, the dots $\dots$ denote terms which are sixth-order and higher in spatial derivatives, $\chi_\perp$ is the transverse magnetic susceptibility, and $\rho$ is the local spin stiffness. Note that since the theory is written in terms of the staggered magnetization, all ordering wavevectors will implicitly be measured from $(\pi,\pi)$. With the aim of applying the large-$N$ technique, we will generalize $\Vec{n} = (n_1,\dots,n_N)$ to have $N \geq 3$ components from hereon. Within linear spin-wave theory, $\rho \sim S^2(J_1 - 4J_3)$. Therefore, at the classical LP $J_3=J_1/4$, we have $\rho = 0$, as well as $b_1 = J_3 S^2 a^2$ and $b_2 = 0$, where $a$ is the lattice spacing of the square lattice; the full expressions for the spin-wave parameters are given in Appendix \ref{app:spinwave}. The N\'eel phase corresponds to $\rho > 0$, while the spiral phase corresponds to $\rho < 0$ and the higher derivative terms in the Lagrangian are necessary to stabilize the spiral ground state; in the rest of this work, we will view the frustration as being tuned by $\rho$. The $(\pm Q,Q)$ spiral order of the $J_1$-$J_3$ model is realized for $b_1 > b_2 \geq 0$. Crucially, the ordering wavevector $Q$ vanishes continuously at the LP, which is why a continuum description is applicable (unlike in the pure $J_1$-$J_2$ model where the transition is from N\'eel to stripe order). We will now include quantum fluctuations and describe their consequences on the classical transition.

\subsection{Destruction of Long-Range Order at \texorpdfstring{$T=0$}{Zero Temperature} \label{sec:RG}}

As was pointed out by Ioffe and Larkin \cite{Ioffe1988}, since the local spin stiffness vanishes at the LP, quantum fluctuations of the collinear or helical staggered magnetization are greatly enhanced. In fact, the fluctuations are so strong that long-range magnetic order is completely destroyed, and an intermediate quantum disordered phase opens up.\footnote{Note that the existence of an \textit{extended} intermediate phase does not hold when the spin-rotation symmetry is only easy plane \cite{Ardonne2004}.} Since the argument of Ioffe and Larkin is instructive and provides a physically transparent interpretation for the formation of a gap to magnetic excitations, we adapt it here. 

First, we define dimensionless couplings $u \equiv \Lambda^{2-z} \chi_\perp^{-1}$, $\varrho \equiv\Lambda^{2-2z} \rho \chi_\perp^{-1}$, $\beta_{1,2} \equiv \Lambda^{4-2z} b_{1,2} \chi_\perp^{-1}$, where $z$ is, for now, an unspecified dynamical exponent denoting the relative scaling of space and time dimensions, and $\Lambda \sim \pi/a$ is the ultraviolet momentum cutoff scale. In the N\'eel phase far from the LP, where $\varrho \gtrsim \beta_{1,2}$, the fast Goldstone modes with momentum $\abs{\vb{q}} \sim \Lambda $ have a dispersion that scales as $\omega_{\vb{q}} \simeq \sqrt{\rho/\chi_\perp} \abs{\vb{q}}$. Therefore, their dynamical exponent is $z=1$, and one recovers the conventional NLSM scaling behavior. In the vicinity of the LP, where $\abs{\varrho} \ll \beta_{1,2} $, the fast Goldstone modes have a dispersion scaling as $\omega_{\vb{q}} \simeq \sqrt{b_{1,2}/\chi_\perp} \vb{q}^2$, and hence, their dynamical exponent is $z=2$. It follows that the crossover between the two regimes occurs at the momentum scale $\Lambda_c \sim \sqrt{\rho/b_{1,2}}$.

Next, one can perform a simple momentum-shell RG calculation similar to that given in Ref. \cite{Ioffe1988} (see Appendix \ref{app:RG} for details). In the $z=2$ regime and to leading order in the couplings, the one-loop RG beta function for $u$ is
\begin{align}
    \frac{d u}{d \ell} =  K\left(\frac{\beta_1 - \beta_2}{2\beta_1}\right) \frac{(N-2) u^2}{2\pi^2 \sqrt{\beta_1}}, \label{eq:kshellbetafunction}
\end{align}
where $\ell = \ln(\Lambda_0/\Lambda)$ is the logarithmic change in the UV cutoff, $K(z)$ is the complete elliptic integral of the first kind, and $\beta_1$, $\beta_2 > 0$ are constant under the RG flow. This beta function is suggestive of the same asymptotic freedom found in the classical 2D NLSM at finite temperature \cite{Chakravarty1989}. Under the same assumption that the flow to strong coupling takes the theory to a quantum disordered phase, there is a dynamically generated energy scale and associated length scale \cite{Polyakov1975,Brezin1976},
\begin{equation}
    \xi \simeq a^{\zeta \sqrt{\beta_1}/u(0)}, \kern1em \zeta^{-1} = \frac{(N-2)}{2\pi^2} K\left(\frac{\beta_1 - \beta_2}{2\beta_1}\right),
\end{equation}
which can be identified with the correlation length of the gapped phase formed by the strong infrared fluctuations.\footnote{This simple scaling argument does not exclude other possibilities such as a direct first order transition between the N\'eel and spiral phases.} However, integrating this beta function down to the scale $\xi$ is only valid if the crossover scale satisfies $\Lambda_c^{-1} < \xi$. Therefore, one can deduce that there is an extended quantum disordered phase in the region $\abs{\rho}/b_{1,2} \lesssim a^{-1} e^{-\zeta \sqrt{\beta_1}/u(0)} \sim a^{-1} e^{-c S}$, where $S$ is the spin in the lattice model and $c \sim \mathcal{O}(1)$ is a dimensionless number. Since $\rho/b_{1,2}$ is a dimension-full quantity that also flows non-trivially under the RG, this is a crude estimate (see Appendix \ref{app:RG} for a detailed discussion of how to more precisely estimate the location of the quantum disordered phase using this RG approach). The key message of this discussion is that, at least within this perturbative RG framework, there is evidence for an extended region in the RG flow where the spin stiffness remains small compared with the correlation length, leading to Goldstone modes with a soft dispersion and strongly enhanced infrared quantum fluctuations. A consistent proof of this can be obtained through a nonperturbative analysis such as the large-$N$ limit, which we present in the next subsection. The rest of this paper is concerned with developing a systematic and nonperturbative method for understanding the nature of this phase using the low energy effective field theory.

\subsection{Large-\texorpdfstring{$N$}{N} Theory of the Quantum \texorpdfstring{$T=0$}{Zero Temperature} Gapped Phase \label{sec:largeNv1}}

Having reviewed the evidence for a quantum disordered phase in the vicinity of the classical LP, in this section we present our own calculation of the phase diagram using the large-$N$ technique; see, for example, Refs. \cite{Das2009,Kharkov2018} for a similar approach. First, we define $\Vec{n} = (\Vec{\pi},\sigma_1,\sigma_2)$. Then, after integrating out only the $N-2$ components of $\Vec{\pi}$, we can write the quantum partition function for our effective field theory Eq. \eqref{eq:lagrangian} in imaginary time $\tau$ at zero temperature $T=0$,
\begin{align}
    \mathcal{Z} &= \int \mathcal{D} \lambda \, \exp\big( - (N-2) \mathcal{S}_{\mathrm{eff}} \big) ,\\
    \mathcal{S}_{\mathrm{eff}} &= \frac{1}{2} \mathrm{tr} \ln \big( - \chi_\perp \del_\tau^2 + \mathcal{K}(\del) + \lambda \big) \nonumber \\
     + \frac{1}{2} &\int_{\vb{x},\tau} \left[ \sum_{\ell=1}^2 \sigma_\ell \left( -\chi_\perp \del_\tau^2 + \mathcal{K}(\del) + \lambda \right) \sigma_\ell - \lambda \right] ,\nonumber \\
    \mathcal{K}(\del) &= -\rho \nabla^2 + b_{ij} \del_i^2 \del_j^2, \nonumber
\end{align}
where $\lambda$ is a Lagrange multiplier enforcing the constraint $\Vec{n}^2 = 1$ and we have defined the tensor $b_{ij}$ with components $b_{xx} = b_{yy} = b_1$ and $b_{xy} = b_{yx} = b_2$. In the $N\to\infty$ limit, the ground state is determined by the solutions to the saddle-point equations
\begin{subequations}
\begin{align}
    \int_{\vb{q},\omega} \frac{1}{\chi_\perp \omega^2 + \mathcal{K}(\vb{q}) + \lambda} + \sigma_1^2 + \sigma_2^2 &= 1 , \label{eq:gap_quantum} \\
    \big(\mathcal{K}(\del) + \lambda \big) \sigma_i &= 0,
\end{align}
\end{subequations}
where the integration measure is the usual $d^2\vb{q} d\omega/(2\pi)^3$, and we have assumed that the solutions are constant in imaginary time; that is, $\del_\tau \sigma_i = 0$. The second equation is a linear eigenvalue equation, so we must also restrict our attention to the set of eigenfunctions with minimal eigenvalues. For $\rho > 0$, the minimum eigenvalue of the differential operator $\mathcal{K}(\del)$ is $\mathcal{K}(0) = 0$, and so the two solutions in this regime are $\lambda = 0$, $\sigma_i \neq 0$, corresponding to the N\'eel phase, and $\lambda > 0$, $\sigma_i = 0$, corresponding to a quantum disordered phase. When $\rho < 0$, $\mathcal{K}(\del)$ has two degenerate minima $\mathcal{K}(\vb{Q}_\pm)$ corresponding to two spiral wavevectors, the direction of which depends on the relative anisotropy $b_1/b_2$,
\begin{subequations} \label{eq:spiralQ}
\begin{align}
    &\vb{Q}_\pm = \sqrt{\frac{\abs{\rho}}{2(b_1 + b_2)}} \big(  \hat{\vb{x}} \pm  \hat{\vb{y}} \big) &\kern1em &\text{if $b_1 > b_2 \geq 0$}, \\
    &\vb{Q}_{x/y} =  \sqrt{\frac{\abs{\rho}}{2b_1}} \hat{\vb{x}}/\hat{\vb{y}} & &\text{if $b_2 > b_1 > 0$} .
\end{align}
\end{subequations}
Note that $\vb{Q}$ and $-\vb{Q}$ represent the same ground state. From hereon we will focus on the case $b_1 > b_2 \geq 0$. Therefore, the solutions in this regime are $\lambda > -\mathcal{K}(\vb{Q}_\pm) > 0$, $\sigma_i = 0$, or $\lambda = -\mathcal{K}(\vb{Q}_\pm)$,
\begin{align}
    \sigma_i(\vb{x}) = c_1 \cos(\vb{Q}_+\cdot\vb{x}) + c_2 \sin(\vb{Q}_+\cdot\vb{x}) \\
    + c_3 \cos(\vb{Q}_-\cdot\vb{x}) + c_4 \sin(\vb{Q}_-\cdot\vb{x}) , \nonumber
\end{align}
where the $c_i$ are as-yet undetermined coefficients. However, since $\lambda$ is a constant, the first saddle-point equation implies that the combination $\sigma_1^2 + \sigma_2^2$ must be uniform in space. Therefore, the actual solutions can only be unidirectional spirals of the form
\begin{equation}
    \sigma_{1}(\vb{x}) = \sigma \cos(\vb{Q}_\pm\cdot\vb{x}), \kern1em \sigma_{2}(\vb{x}) = \sigma \sin(\vb{Q}_\pm\cdot\vb{x}),
\end{equation}
for some constant amplitude $\sigma$ and up to an arbitrary internal $O(2)$ transformation of $(\sigma_1,\sigma_2)$. We note that the restriction to a unidirectional spiral was a consequence of looking for modulation in only two components $\sigma_1$ and $\sigma_2$ of $\Vec{n}$. However, since our aim is to make contact with the limit $N = 3$, this constraint is physically motivated.

Since the integral in Eq. \eqref{eq:gap_quantum} is ultraviolet divergent, we first need to apply a renormalization procedure: Performing the frequency integral and defining $g = \chi_\perp^{-1/2}$, we can re-write the saddle-point equation as
\begin{equation}
    \int_{\vb{q}} \frac{1}{2 \sqrt{ \mathcal{K}(\vb{q}) + \lambda}} + \frac{\sigma_1^2 + \sigma_2^2 - 1}{g} = 0.
\end{equation}
Then, it suffices to define a renormalized coupling $g_R$
\begin{equation}
    \frac{1}{g_R} = \frac{1}{g} - \int_{\vb{q}} \frac{1}{2 \sqrt{ \smash[b]{b_{ij}q_i^2 q_j^2 + \mu^2}}} ,
\end{equation}
where $\mu$ is the renormalization scale, and the fields $\sigma_{i}$ such that $\sigma_{i}^2/g = \sigma_{i,R}^2/g_R$. Note that this expression for the renormalization of $g$ implies the RG beta function,
\begin{equation}
    -\mu\frac{\del g_R}{\del \mu} = K\left(\frac{b_1 - b_2}{2b_1}\right) \frac{g_R^2}{4\pi^2 \sqrt{b_1}},
\end{equation}
which displays the same asymptotic freedom as in the perturbative momentum shell RG analysis discussed above; note that $K(z)$ here is again the elliptic integral of the first kind.

Next, we can determine the phase boundaries approximately analytically as a function of $\rho$ by looking for solutions to the saddle-point equation \eqref{eq:gap_quantum} where $\sigma_i = 0$ and $\lambda = 0$ (for the N\'eel to quantum disordered critical point) and $\lambda = -\mathcal{K}(\vb{Q}_\pm)$ (for the spiral to quantum disordered critical point). First, for the case $\rho > 0$, we can evaluate the saddle-point equation to logarithmic accuracy to find the N\'eel to quantum disordered transition point
\begin{equation}
    \rho_{cN} \simeq 2\mu \sqrt{\bar{b}} e^{-\zeta \sqrt{b_1} /g_R},
\end{equation}
where $\bar{b}$ is some suitably defined average of $b_\theta = b_1(\cos^4\theta+\sin^4\theta) + 2b_2 \cos^2\theta\sin^2\theta$ over $\theta\in[0,2\pi)$, for example, the geometric mean $\sqrt{b_0 b_{\pi/4}}$, and we have defined
\begin{equation}
    \zeta^{-1} = \frac{1}{4\pi^2} K\left(\frac{b_1 - b_2}{2b_1}\right).
\end{equation}
Observe that the essential singularity in $g_R/\sqrt{b_1}$ is qualitatively consistent with the perturbative RG discussed above. Then, for the case $\rho < 0$ we similarly find the spiral to quantum disordered transition point to be given by
\begin{equation}
    \rho_{cS} \simeq -2\mu \sqrt{\bar{b}}  \left(\sqrt{\frac{\Bar{b}}{b_{\vb{Q}}}} + 1\right) \left(\frac{8 b_{\vb{Q}}}{\abs{b_1 - b_2}}\right)^{\kappa} e^{-\zeta\sqrt{b_1}/g_R},
\end{equation}
where $b_{\vb{Q}} = (b_1+b_2)/2$ if $b_1\geq b_2$ and $b_{\vb{Q}}= b_1$ if $b_2>b_1$, and we have defined the non-universal power law exponent
\begin{equation}
    \kappa^{-1} = \frac{2}{\pi} \sqrt{\frac{b_{\vb{Q}}}{b_1}} K\left(\frac{b_1 - b_2}{2b_1}\right),
\end{equation}
where $\kappa \geq 1$ and the lower bound is only attained when $b_2 = b_1$. Therefore, in the isotropic limit $b_2 \to b_1$, the spiral phase is pushed to $\rho \to -\infty$ since the continuum of degenerate ordering wavevectors $\vb{Q}$ leads to a new Goldstone mode, destabilizing the ordered phase; note that in this scenario effects beyond the current analysis can stabilize new ordered phases through the order-by-disorder mechanism (see, for instance, Refs. \cite{Bergman2007,Hsieh2022}). 

In both cases, the amplitude of the magnetic order parameters has the form
\begin{equation}
    \sigma_{1,R}^2 + \sigma_{2,R}^2 =
        \frac{g_R}{\zeta \sqrt{b_1}} \ln\left(\abs{\frac{\rho}{\rho_{c}}}\right),
\end{equation}
where $\rho_c = \rho_{cN}$ for $\rho > \rho_{cN}$ and $\rho_c = \rho_{cS}$ for $\rho < \rho_{cS}$, implying that both transitions are continuous at $N=\infty$.

Finally, we consider the correlations of the staggered magnetization fluctuations in the quantum disordered phase. The (renormalized) dynamic susceptibility at $N=\infty$ is simply the analytic continuation back to real frequency of the $\vec{n}$ field propagator
\begin{equation}
    \chi_n(\vb{q},\omega) = \frac{g_R}{(\omega + i0^+)^2 - \mathcal{K}(\vb{q}) - \lambda} ,
\end{equation}
while the static structure factor is
\begin{equation}
    S(\vb{q}) = \frac{g_R}{2\sqrt{\mathcal{K}(\vb{q}) + \lambda}}.
\end{equation}
Therefore, for $\rho \geq 0$ the response is peaked at $\vb{q} = 0$, while for $\rho < 0$ the response has equally-weighted peaks at the four wavevectors $\vb{q} = \pm \vb{Q}_\pm$, signaling short-range incommensurate fluctuations. The correlation length is controlled by the gap in the magnon dispersion: $m^2 = \lambda$ for $\rho \geq 0$ and $m^2 = \lambda - \rho^2/4b_{\vb{Q}}$ for $\rho < 0$. As discussed in the previous subsection, proximity to the LP $\rho = 0$ controls the relative scaling between length and energy, so the correlation length will crossover between $\xi\sim m^{-1}$ ($z=1$) near the ordering transitions and $\xi\sim m^{-1/2}$ ($z=2$) near the LP. Precisely at $\rho = 0$, the saddle-point equation can be solved exactly to yield
\begin{equation}
    m = \mu e^{-\zeta \sqrt{b_1}/g_R} ,
\end{equation}
consistent with the estimate from perturbation theory. The gap as a function of general $\rho_{cS} < \rho < \rho_{cN}$ can only be determined numerically and is discussed in greater detail in Sec. \ref{sec:quantumvestigial}.

Having demonstrated the existence of the quantum disordered phase using the nonperturbative large-$N$ technique, we are left with two guiding questions: 
\begin{enumerate}[label=(\roman*),leftmargin=0pt,itemindent=25pt]
    \item Contributions to the action which are fourth order in spatial derivatives are marginal at the classical LP, including higher-order interactions such as $(\del_i \Vec{n})^4$, $(\del_x \Vec{n} \cdot \del_y \Vec{n})^2$, or specifically in the case of $N=3$, $(\epsilon_{abc} n^a \del_x n^b \del_y n^c)^2$. How do the interactions we have neglected so far influence the phase diagram?
    \item The classical phases of the 2D $J_1$-$J_3$ model are well-understood, including the existence of an Ising nematic phase. How do phases with broken symmetries at finite temperature evolve down to the quantum regime at zero temperature?
\end{enumerate}

\section{Understanding the Classical Phase Diagram \label{sec:classical}}

A useful starting point to build our intuition will be to consider another, simpler and better-understood, regime in which the system is known to exhibit a gap to magnetic excitations: finite temperature. In this section, we will start by reviewing some prior work and then use it as a foundation to introduce the main technique of this work. Along the way, we will demonstrate our first key result; effective field theory and the large-$N$ technique can be used to describe the vestigial Ising nematic phase of the classical $J_1$-$J_3$ model.

The classical phase diagram of the $J_1$-$J_3$ model as a function of temperature has been well-understood for some time \cite{Capriotti2004}. Due to the Mermin-Wagner theorem, long-range N\'eel and spiral order are destroyed by thermal fluctuations. However, the situation is more subtle on the spiral side of the LP. As noted above, there are two degenerate wavevectors $\vb{Q}_\pm = (\pm Q,Q)/\sqrt{2}$. These two configurations can only be related by a combination of global spin rotation and reflection about the $x$ or $y$ axes. Therefore, the internal $O(3)$ symmetry mixes with the $C_{4v}$ point group of the square lattice to yield an order parameter manifold $O(3) \times \mathbb{Z}_2$. Spirals with distinct $\mathbb{Z}_2$ symmetry breaking can be distinguished on the lattice by the plaquette order parameter
\begin{equation}
    \Psi_i = \vb{S}_{i}\cdot \vb{S}_{i+\hat{\vb{x}}+\hat{\vb{y}}} - \vb{S}_{i+\hat{\vb{x}}} \cdot \vb{S}_{i+\hat{\vb{y}}}. \label{eq:isinglattice}
\end{equation}
Crucially, the Mermin-Wagner theorem says nothing about the discrete Ising $\mathbb{Z}_2$ subgroup of the global symmetry, which is not forbidden from being spontaneously broken at finite temperature. Physically, one can visualize that the unidirectional spin spiral that exists at zero temperature is melted by thermal fluctuations, while the system retains knowledge of the broken spatial rotation/mirror symmetry through short-range spiral correlations; this is the quintessential behavior of fluctuating stripe phases with vestigial nematic order. The existence of an Ising nematic phase at finite temperature was demonstrated numerically using Monte Carlo techniques in Ref. \cite{Capriotti2004}. In that work, the authors also applied the large-$N$ technique to the low energy effective theory in Eq. \eqref{eq:lagrangian}. In the continuum, the analog of the nematic order parameter Eq. \eqref{eq:isinglattice} is
\begin{equation}
    \Psi = \Vec{n} \cdot \del_x \del_y \Vec{n} - \del_x \Vec{n}\cdot \del_y \Vec{n},
\end{equation}
and within the low energy effective theory, the corresponding susceptibility in the $N\to\infty$ limit is
\begin{equation}
    \chi_\Psi = T \int_{\vb{q}} \frac{q_x^2 q_y^2}{[\rho \vb{q}^2 + b_{ij} q_i^2 q_j^2 + \lambda]^2}, \label{eq:susceptbare}
\end{equation}
where $\lambda > 0$ is a Lagrange multiplier that enforces the unit vector constraint $\Vec{n}^2 = 1$, and is determined self-consistently by the gap equation
\begin{equation}
    T \int_{\vb{q}} \frac{1}{\rho \vb{q}^2 + b_{ij} q_i^2 q_j^2 + \lambda} = 1.
\end{equation}
As noted in Ref. \cite{Capriotti2004}, it is straightforward to check that the susceptibility in Eq. \eqref{eq:susceptbare} only diverges at zero temperature and $\rho < 0$; i.e., only in the zero temperature spiral ground state and not at some finite temperature nematic transition, appearing to contradict the Monte Carlo data. In the rest of this section, we will show that this apparent contradiction is resolved by including higher-order interactions which have previously been neglected.

\subsection{Effective Field Theory} \label{sec:effectivefieldtheory}

The apparent disagreement between Mote Carlo simulations and the large-$N$ technique presented in Ref. \cite{Capriotti2004} is key to understanding our own approach in this work and its resolution will be our first main result. Within the framework of effective field theory, one must include all possible terms in the action with the same scaling dimensions that are allowed by symmetry. In the present case, since the terms in Eq. \eqref{eq:lagrangian} which are fourth order in derivatives were necessary to stabilize the spiral phase, they must also be accompanied by (technically irrelevant) four-derivative operators of the form
\begin{align}
    \mathcal{L}_{\mathrm{int}} = \frac{\gamma_0}{8} [(\grad \Vec{n})^2]^2 + \frac{\gamma_1}{8} \left[ (\del_x \Vec{n})^2 - (\del_y \Vec{n})^2 \right]^2 \label{eq:intLagrangian} \\
    + \frac{\gamma_2}{2} (\del_x \Vec{n} \cdot \del_y \Vec{n})^2 , \nonumber
\end{align}
where stability requires $\gamma_0 + \gamma_1 + \gamma_2 > 0$; specifically, we will need $\gamma_0 > 0$ to access the regime where $\gamma_1$ or $\gamma_2< 0$. We may add the additional requirement of having a well-defined large-$N$ limit, excluding terms such as $(\epsilon_{abc} n^a \del_x n^b \del_y n^c)^2$. While it is true that the coefficients $\gamma_i$ of such terms vanish in a linear spin-wave expansion, they will necessarily be generated in any calculation of nonlinear spin dynamics. Additionally, while one may argue that all contributions which are fourth order in derivatives are RG irrelevant at finite temperature, irrelevant operators that break symmetries (so-called dangerously irrelevant operators) can play an important role in determining ground state properties of a theory \cite{Fisher-1974,Bruce-1975,Amit-1982,Millis-1993,Ghaemi2005}. Indeed, these terms can and will have a profound effect on the calculation of the nematic susceptibility, as we will see below. The interactions in Eq. \eqref{eq:intLagrangian} can be included in a large-$N$ calculation by using Hubbard-Stratonovich transformations to decouple each term into contributions which are only quadratic in $\Vec{n}$. After such a transformation, the $N$ components of $\Vec{n}$ can be integrated out and the partition function for the theory can be written in the form
\begin{align} \label{eq:classical_partition}
    \mathcal{Z}[H_1,H_2] &= \int \mathcal{D} \lambda \mathcal{D} \eta \mathcal{D} \psi_1 \mathcal{D} \psi_2 \, \exp\big( - N \mathcal{S}_{\mathrm{eff}} \big) ,\\
    \mathcal{S}_{\mathrm{eff}} &= \frac{1}{2} \mathrm{tr} \ln \big( \mathcal{K}(\del,\rho[H]) + \lambda \big) \nonumber \\
    &- \frac{1}{2 T} \int_{\vb{x}} \left( \lambda + \frac{\eta^2}{\gamma_0} + \frac{\psi_1^2}{\gamma_1} + \frac{\psi_2^2}{\gamma_2}\right), \nonumber \\
    \mathcal{K}(\del,\rho[H]) &= - \rho_{ij}[H] \del_i\del_j + b_{ij} \del_i^2 \del_j^2 ,\nonumber\\
    \rho_{ij}[H] &= \begin{pmatrix}
        \rho + \eta + \psi_1  + H_1 & \psi_{2} + H_2 \\
        \psi_{2} + H_2 & \rho + \eta - \psi_1 - H_1
    \end{pmatrix}_{ij}, \nonumber
\end{align}
where $T$ is the temperature, $\lambda$ enforces the unit vector constraint $\Vec{n}^2 = 1$, $\eta$, $\psi_1$ and $\psi_2$ are Hubbard-Stratonovich fields which decouple each term in Eq. \eqref{eq:intLagrangian}, and
$H_1$ and $H_2$ are external symmetry-breaking sources which couple to the two distinct nematic components allowed by the $C_{4v}$ symmetry of the square lattice:
\begin{equation}
    \mathcal{L}_{\mathrm{sources}} = H_1 \Vec{n} \cdot (\del_x^2 - \del_y^2)\Vec{n} + H_2 (\Vec{n} \cdot \del_x \del_y \Vec{n} - \del_x \Vec{n}\cdot \del_y \Vec{n}) .
\end{equation}
Instead of the sources $H_1$ and $H_2$, it will be more instructive to work in terms of the nematic order parameters
\begin{subequations} 
\begin{align}
    \Psi_1 = \langle \Vec{n}\cdot(\del_x^2 - \del_y^2)\Vec{n} \rangle, \\
    \Psi_2 = \langle \Vec{n} \cdot \del_x \del_y \Vec{n} - \del_x \Vec{n}\cdot \del_y \Vec{n}\rangle . 
\end{align}
\end{subequations}
It follows from the partition function that
\begin{align} \label{eq:nematicOPs}
    \Psi_i \equiv - \frac{\delta \mathcal{F}[H_1,H_2]}{\delta H_i} = \frac{H_i - \langle\psi_i\rangle_H}{\gamma_i},
\end{align}
where $\mathcal{F}[H_1,H_2] = -T \ln\mathcal{Z}[H_1,H_2]$ is the free energy and $\langle \psi_i \rangle_H$ is the thermodynamic average of $\psi_i$ in the presence of the external sources. Then, the effective potential as a function of $\Psi_1$ and $\Psi_2$ is obtained by a Legendre transform:
\begin{equation}
    \Gamma[\Psi_1,\Psi_2] = H_1 \Psi_1 + H_2 \Psi_2 + \mathcal{F}[H_1,H_2],
    \label{eq:fh1h2}
\end{equation}
which satisfies
\begin{equation}
    \frac{\delta \Gamma[\Psi_1,\Psi_2]}{\delta \Psi_i} = H_i, \label{eq:effpotsources}
\end{equation}
and in the $N\to\infty$ limit, we have
\begin{flalign} 
    &\Gamma[\Psi_1,\Psi_2] =  - \frac{\lambda}{2} - \frac{(\eta - \rho)^2}{2\gamma_0} \label{eq:classEffPot1} &&\\
    &+\frac{\gamma_1 \Psi_1^2}{2} + \frac{\gamma_2 \Psi_2^2}{2}  + \Psi_1 \psi_1 + \Psi_2 \psi_2  \nonumber &&\\
    &+\frac{T}{4} \int_{\vb{q}} \ln \left( [\eta \vb{q}^2 + b_{ij} q_i^2 q_j^2 + \lambda]^2 - [\psi_{ij}q_iq_j]^2 \right)  , \nonumber
\end{flalign}
where $\psi_{xx}=-\psi_{yy}=\psi_1$ and $\psi_{xy}=\psi_{yx}=\psi_2$. Here $\eta$ enters as the uniform component of the physical local spin stiffness tensor renormalized by the quartic interactions, while the nematicity enters through the traceless components of the stiffness tensor $\psi_1$ and $\psi_2$, which appear in the effective action in a $C_{4v}$ symmetric way. 

The details of evaluating the potential are cumbersome and not instructive, so they are given in Appendix \ref{app:saddles}. However, we emphasize that the theory at $N=\infty$ is entirely characterized by the solutions to the four coupled saddle-point equations
\begin{equation}
    \frac{\delta \Gamma}{\delta \lambda} = \frac{\delta \Gamma}{\delta \eta} = \frac{\delta \Gamma}{\delta \psi_1} = \frac{\delta \Gamma}{\delta \psi_2} = 0,
\end{equation}
and note that the solution requires renormalization of the stiffness parameter $\rho \to \rho_R$; the (nonlinear) relation between the tuning parameter $\rho_R$ and the physical local stiffness $\eta$ is obtained by solving the saddle-point equations.

Finally, we note that the $N=\infty$ static structure factor for the thermodynamic fluctuations of the staggered magnetization can be read off directly from the effective action
\begin{equation}
    S(\vb{q}) = \frac{T}{(\eta\delta_{ij} + \psi_{ij})q_iq_j + b_{ij} q_i^2 q_j^2 + \lambda } .
\end{equation}
Therefore, we see that in addition to diagnosing the crossover from commensurate to incommensurate fluctuations as noted in the previous section, the structure factor also reveals nematicity through asymmetry in the spectral weight at each of the spiral peaks, which occurs when either $\psi_1$ or $\psi_2 \neq 0$. Next, we will develop our technique for understanding the physics encoded in the large-$N$ effective action.

\begin{figure}
    \centering
    \includegraphics[width=0.85\linewidth]{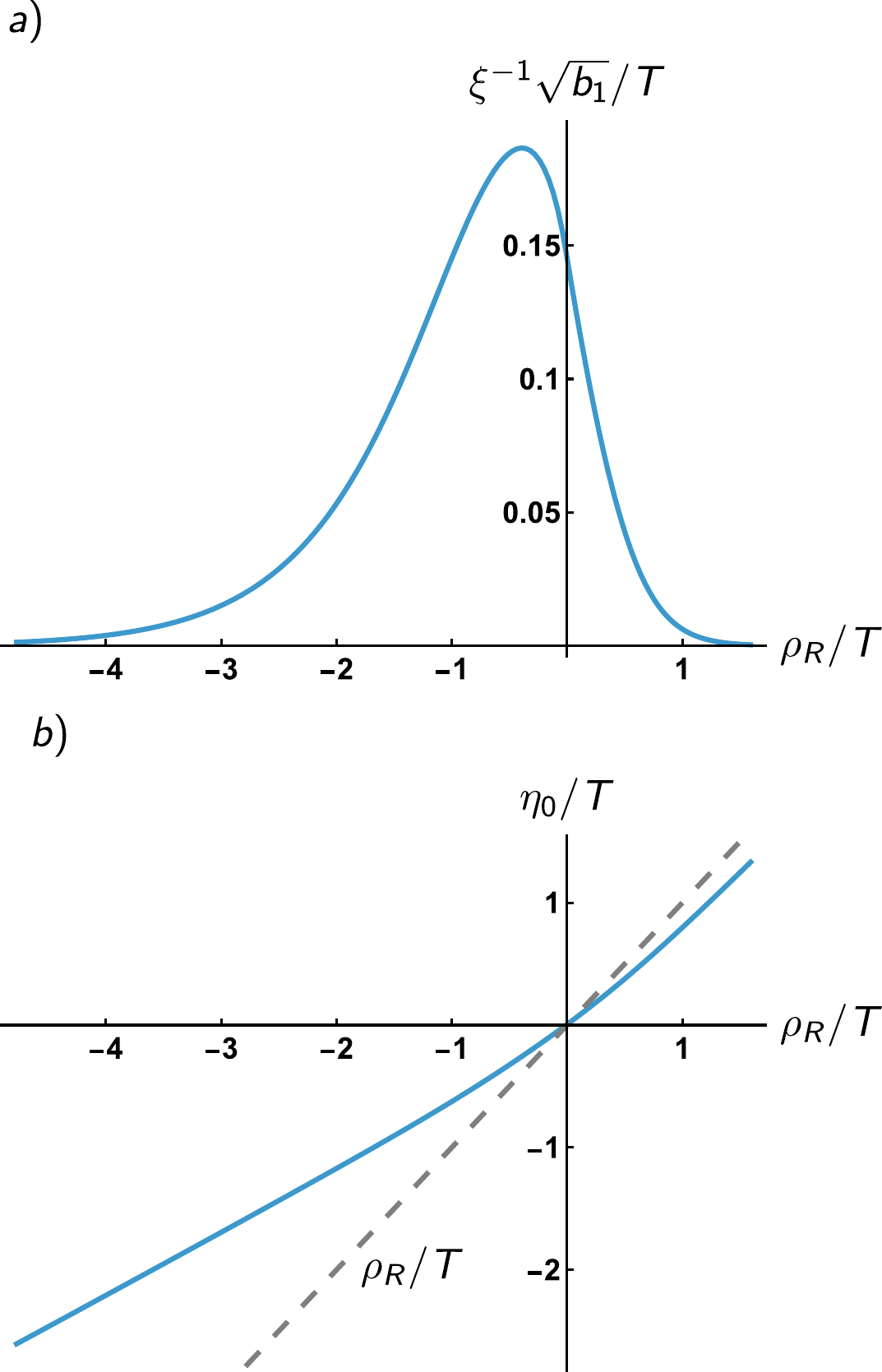}
    \caption{(a) The dimensionless inverse correlation length $\xi^{-1}\sqrt{b_1}/T$, where $\xi^{-2} = \lambda_0$ for $\eta_0 > 0$ and $\xi^{-2} = \lambda_0 -\eta_0^2/4b_{\vb{Q}}$ for $\eta_0 < 0$, and (b) the dimensionless physical stiffness $\eta_0/T$, both as a function of the renormalized stiffness parameter $\rho_R/T$, for $b_2 = 0$, $\mu^2b_1/T^2 = 1$ and $\gamma_0/b_1 = 2$. In (b), the physical stiffness, which is modified by higher derivative interactions, is compared with the renormalized stiffness parameter shown by the dashed gray line, which is the stiffness of the theory in the absence of higher derivative interactions.}
    \label{fig:classical_saddle_sols}
\end{figure}

\subsection{Ginzburg-Landau Expansion}

The saddle-point equations imply that their solutions $\lambda$, $\eta$, $\psi_1$ and $\psi_2$ are implicit functions of the order parameters $\Psi_1$ and $\Psi_2$, which enter as free variables in the effective potential $\Gamma[\Psi_1,\Psi_2]$. Since these saddle-point equations can only be solved numerically, to gain analytical insight it is useful to expand the effective potential Eq. \eqref{eq:classEffPot1} in a Ginzburg-Landau expansion
\begin{equation} \label{eq:ginzburg-landau}
    \Gamma[\Psi_1,\Psi_2] = \sum_{\ell,m=0}^\infty \frac{\Gamma^{(2\ell,2m)}[0]}{(2\ell)! (2m)!}  \Psi_1^{2\ell} \Psi_2^{2m},
\end{equation}
where the second order coefficients $\Gamma^{(2,0)}[0]$ and $\Gamma^{(0,2)}[0]$ are the inverses of the nematic susceptibilities in the absence of long-range nematic order, and only even powers of $\Psi_1$ and $\Psi_2$ enter due to the $C_{4v}$ symmetry.

To simplify the following analysis, we use the observation that when $b_1 > b_2$, the $T=0$ ground state is a $(\pm Q,Q)$ spiral which has the symmetry of the $\Psi_2$ nematic order parameter. Therefore, it is clear that even for comparable interaction strengths $\gamma_1$ and $\gamma_2$, the $\Psi_1$ nematic will be energetically suppressed in our regime of interest, and hence, we will set $\Psi_1 \equiv 0$ and $\Psi_2 \equiv \Psi$ from hereon; this assumption clearly breaks down if $\gamma_1$ is substantially more negative than $\gamma_2$. It is simple to check that this implies $\psi_1 = 0$ is the only solution of the saddle-point equations. Then, expanding in powers of $\Psi$, noting that only $\psi(\Psi)$ is odd under the Ising nematic symmetry,
\begin{subequations}
\begin{align}
    \lambda(\Psi) &= \lambda(0) + \frac{1}{2!}\lambda^{(2)}(0)\Psi^2  + \dots, \\
    \eta(\Psi) &= \eta(0) + \frac{1}{2!}\eta^{(2)}(0)\Psi^2 + \dots, \\
    \psi(\Psi) &= \psi^{(1)}(0) \Psi + \frac{1}{3!}\psi^{(3)}(0) \Psi^3 +  \dots,
\end{align}
\end{subequations}
and substituting into the saddle-point equations, we solve for each of the expansion coefficients $\lambda^{(i)}$, $\eta^{(i)}$ and $\psi^{(i)}$ order by order.

First, we note some properties of the theory when $\Psi = 0$. The $\Psi=0$ parameters $\lambda_0 = \lambda(0)$ and $\eta_0 = \eta(0)$ can be determined numerically exactly as functions of $\rho_R$ from the saddle-point equations. The physical local stiffness $\eta_0$ differs little from $\rho_R$, as can be seen in Fig. \ref{fig:classical_saddle_sols} (b), particularly at low temperatures, while the Lagrange multiplier $\lambda_0$ has the following asymptotic behavior:
\begin{equation} \label{eq:lambda_asymptotic}
    \lambda_0 = \begin{cases}
        \dfrac{8 \eta_0^2 e^{-\frac{4\pi \eta_0}{T}} }{3b_1 + b_2 + 2\sqrt{2b_1(b_1+b_2)}}, \kern0.5em & \eta_0 \gg T, \\[1em]
        \dfrac{T^2}{b_1}\left[ \dfrac{1}{4\pi} K\left(\dfrac{b_1 - b_2}{2b_1}\right) \right]^2, \kern0.5em & \eta_0 = 0, \\[1em]
        \dfrac{\eta_0^2}{4 b_{\vb{Q}}}\left(1 + c e^{-\frac{8\pi \abs{\eta_0}}{T}\sqrt{\frac{b_1 - b_2}{b_1+b_2}}} \right), \kern0.5em &   \eta_0 <0, \vert\eta_0\vert \gg T,
    \end{cases}
\end{equation}
where the last case is determined only to logarithmic accuracy (i.e., the exponent but not the coefficient) and $c\sim \mathcal{O}(1)$ is a dimensionless number. When $\eta_0 > 0$, the correlation length is $\xi = \lambda_0^{-1/2}$ and is isotropic in space, whereas in the regime with spiral order at zero temperature ($\eta_0 < 0$) the longest correlation length is along the $\vb{Q}_\pm$ directions, and has the form $\xi^{-2} = \lambda_0 -\eta_0^2/4b_{\vb{Q}}$. The inverse correlation length is shown in Fig. \ref{fig:classical_saddle_sols} (a); note the two distinct exponential tails for $\eta_0>0$ and $\eta_0 < 0$, in agreement with the asymptotic behavior given in Eq. \eqref{eq:lambda_asymptotic}. Therefore, away from $\eta_0 = 0$, the inverse of the characteristic correlation length $\xi^{-1}$ is always exponentially small at low temperatures. When $\eta_0 < 0$, this signals fluctuating spirals with an exponentially large correlation length along the two diagonals $\vb{Q}_\pm$. The very large correlation length in particular implies that the system is very close to magnetically ordering and hence, that there is a possibility of nematic order.

\begin{figure}
    \centering
    \includegraphics[width=\linewidth]{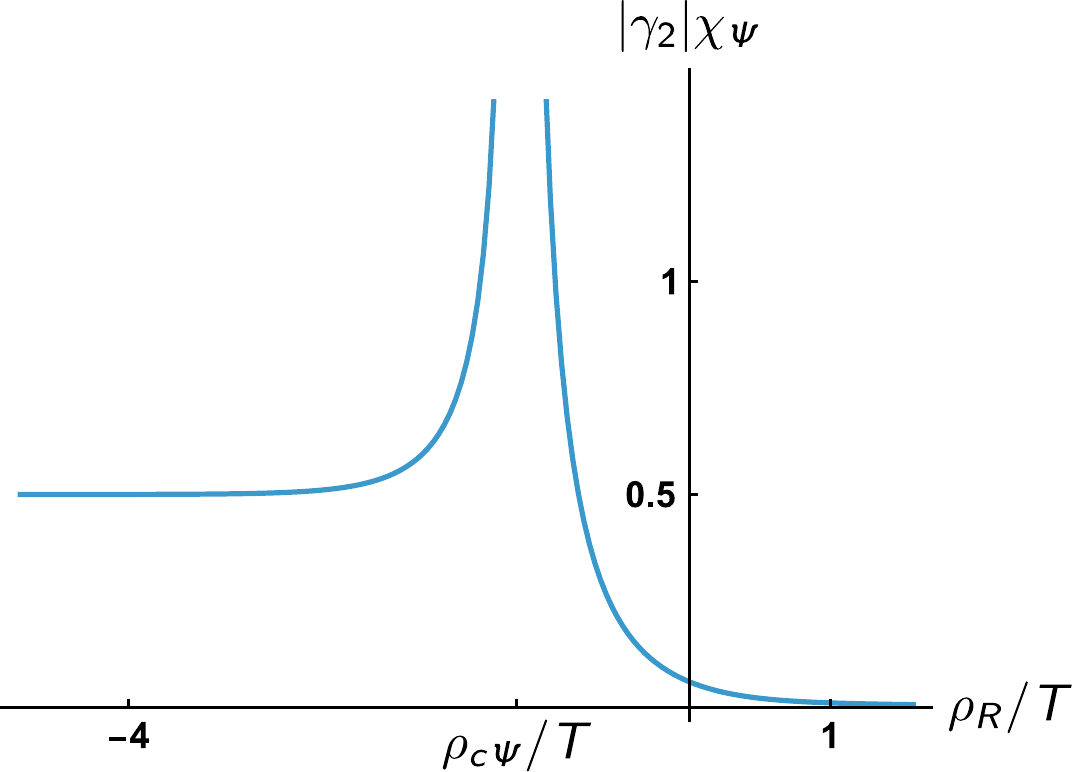}
    \caption{Classical nematic susceptibility as a function of $\rho_R/T$ for $b_2 = 0$, $\mu^2b_1/T = 1$, $\gamma_0/b_1 = 2$ and $\gamma_2/\gamma_0 = -1/8$. Here $\rho_{c\Psi}/T \approx -1.2$. For $\rho_R < \rho_{c\Psi}$ the susceptibility in the nematic phase follows from the Ginzburg-Landau analysis as $\chi_\Psi = \vert 2\Gamma^{(0,2)}[0]\vert^{-1}$.}
    \label{fig:classical_suscept}
\end{figure}

Next, by combining Eqs. \eqref{eq:nematicOPs} and \eqref{eq:effpotsources} as
\begin{equation} \label{eq:legendre_identity}
    \frac{\delta \Gamma[\Psi_1,\Psi_2]}{\delta\Psi_i} = \gamma_i \Psi_i + \langle\psi_i\rangle_\Psi,
\end{equation}
where $\langle\psi_i\rangle_\Psi$ should be viewed as an implicit function of the order parameters, we see that the Ginzburg-Landau coefficients defined in Eq. \eqref{eq:ginzburg-landau} satisfy
\begin{equation} \label{eq:GL_identity}
    \Gamma^{{(0,2)}}[0] = \gamma_2 + \psi^{(1)}, \kern1em \Gamma^{{(0,k)}}[0] = \psi^{(k-1)},
\end{equation}
for $k \geq 4$. Note that due to the Ising symmetry, $\psi^{(k-1)}$ depends only on coefficients up to $\lambda^{(k-2)}$ and $\eta^{(k-2)}$. In particular, we find
\begin{align}
    &\Gamma^{{(0,2)}}[0] = \gamma_2 + \frac{1}{T}\Bigg(\int_0^{2\pi} \frac{d\theta}{2\pi} \frac{\sin^2(2\theta)}{8\pi b_\theta}\\
    &\times \left[ \frac{4b_\theta\lambda_0\cos ^{-1}\left(\eta_0/\sqrt{4 b_\theta \lambda_0}\right)}{(4 b_\theta  \lambda_0-\eta_0^2)^{3/2}} - \frac{\eta_0}{4b_\theta \lambda_0 - \eta_0^2} \right]  \Bigg)^{-1} . \nonumber
\end{align}
This yields a nematic susceptibility as a function of $\rho_R$ shown in Fig. \ref{fig:classical_suscept} for $\gamma_2/\gamma_0 = -1/8$. Therefore, given the divergence in the susceptibility at a critical value $\rho_{c\Psi}$, we have shown in this section that despite the quartic interactions in Eq. \eqref{eq:intLagrangian} being RG irrelevant, they are crucial for stabilizing nematic order at finite temperature (that is, they are dangerously irrelevant). In fact, any negative $\gamma_2$ is sufficient to stabilize nematic order at a low enough temperature. This is our first main result: The large-$N$ technique and field theory can in fact describe the vestigial nematic phase of the classical $J_1$-$J_3$ model.

\begin{figure}
    \centering
    \includegraphics[width=\linewidth]{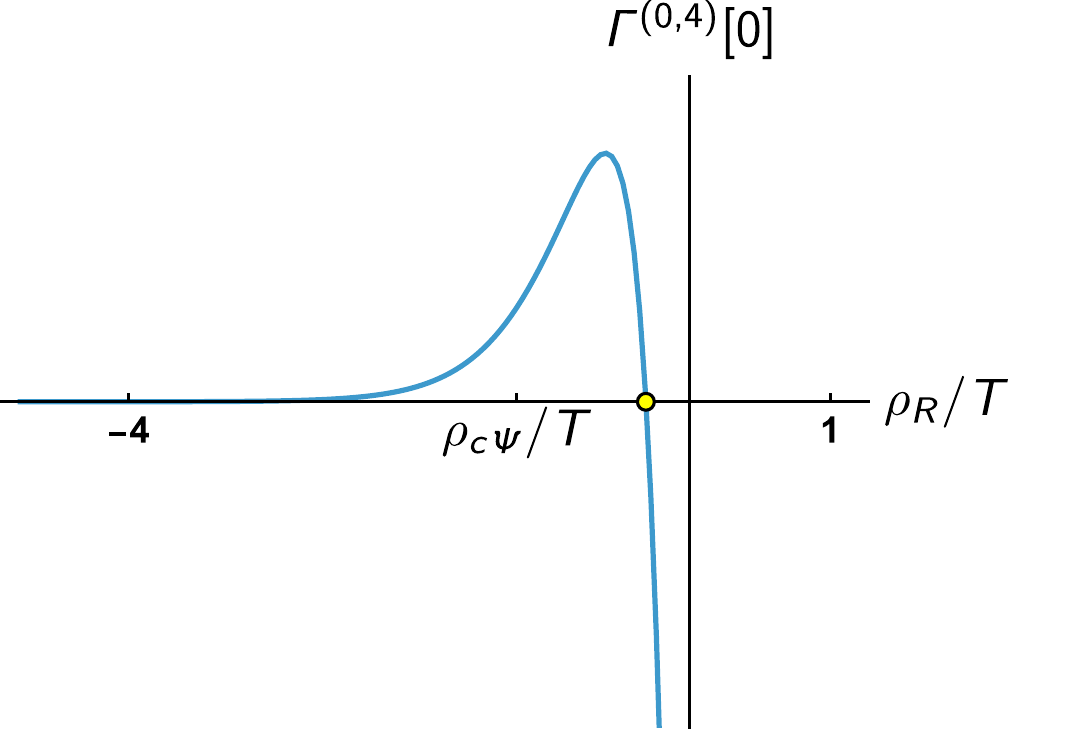}
    \caption{Fourth order Ginzburg-Landau coefficient (arbitrary units) as a function of $\rho_R/T$ for $b_2 = 0$, $\mu^2b_1/T = 1$, $\gamma_0/b_1 = 2$ and $\gamma_2/\gamma_0 = -1/8$. Here $\rho_{c\Psi}/T \approx -1.2$ and the tricritical point marked with the yellow dot is at $\rho_R/T \approx -0.3$.}
    \label{fig:classical_quart}
\end{figure}

Naturally, a putative ordered phase is only stable within the Ginzburg-Landau analysis if higher-order coefficients $\Gamma^{(0,2\ell)}[0]$ are positive. It is straightforward to extend the power series solution to higher orders: At even orders $2\ell$, the saddle-point equations yield a system of two coupled equations which are \textit{linear} in $\lambda^{(2\ell)}(0)$ and $\eta^{(2\ell)}(0)$, and at the next odd order, there is a single linear equation for $\psi^{(2\ell+1)}(0)$ which depends only on previous orders. This way, using Eq. \eqref{eq:GL_identity} we can easily numerically solve for any higher-order coefficient we wish. In particular, $\Gamma^{(0,4)}[0]$ is shown in Fig. \ref{fig:classical_quart}, and has three important features: 
\begin{enumerate}[label=(\roman*),leftmargin=0pt,itemindent=25pt]
    \item It vanishes exponentially as $\rho_R/T \to -\infty$, but remains positive in this limit.
    \item It rapidly becomes negative for $\rho_R \to 0^-$ from below.
    \item For $\gamma_2$ negative but small enough, it remains positive at the critical point $\rho_{c\Psi}$.
\end{enumerate}
Therefore, for small $\gamma_2$, the Ginzburg-Landau analysis implies a continuous phase transition. The fourth order coefficient becoming negative as $\rho_R/T$ approaches $0$ is also not cause for concern, since one can easily verify that $\Gamma^{(0,6)}[0]$ is always positive in that regime, guaranteeing boundedness and stability of the power series expansion. However, it is suggestive of, at least in the $N \to \infty$ limit, the existence of a tricritical point between a regime with a continuous nematic transition for small negative values of $\gamma_2$ to a regime with a weakly first-order transition for more negative values of $\gamma_2$, where the corresponding critical coupling $\rho_{c\Psi}$ becomes larger (more positive). Note that this does not necessarily imply that such a regime is accessible within the microscopic $J_1$-$J_3$ model, and to our knowledge no such point has been observed in that model.

Importantly, the above results are valid precisely when $N\to\infty$, where composite orders such as the Ising nematic are treated as uniform mean-field parameters. However, long-range order does not necessarily set in at the same temperature as the formation of the order parameter amplitude; the amplitude can be ``pre-formed''. It is well-understood that the 2D Ising universality class, which describes continuous phase transitions of the Ising nematic order which we study in this work, is characterized by the proliferation of domain walls in the (pre-formed) amplitude. Unfortunately, Ising domain walls are inherently dominated by ultraviolet scale physics that only allows for a qualitative mapping to an effective Ising model description (see, for example, Ref. \cite{Nie2014}). Therefore, determining precise phase boundaries, particularly in regimes where the $N\to\infty$ limit predicts a first-order mean-field transition, requires a complicated analysis and comparison with domain wall energies that is beyond the scope of this work, given that we have shown our method reproduces the vestigial nematic phase, as well as the complete picture of the classical $J_1$-$J_3$ model that already exists \cite{Capriotti2004}.

\section{Quantum Vestigial Nematic Order at \texorpdfstring{$T=0$}{Zero Temperature} \label{sec:quantumvestigial}}

Having developed a large-$N$ theory of the classical vestigial nematic, we are now in a position to apply our technique to the quantum Lifshitz field theory at zero temperature. In this section, we will demonstrate the second key result of this work: Vestigial nematic order can be stabilized out of the fluctuating spirals in the vicinity of the quantum disordered to spiral phase transition.

As in the previous section, we will begin by considering the nematic susceptibility of the quadratic action in the gapped phase
\begin{equation}
    \chi_\Psi = \int_{\vb{q},\omega} \frac{q_x^2 q_y^2}{[\chi_\perp \omega^2 + \rho \vb{q}^2 + b_{ij} q_i^2 q_j^2 + \lambda]^2} ,
\end{equation}
where the Lagrange multiplier $\lambda$ is determined by the saddle-point equation Eq. \eqref{eq:gap_quantum}. This integral is logarithmically ultraviolet divergent. However, in its current form the theory cannot be renormalized in any way to cure this divergence; the nematic correlations are inherently more singular and dominated by shorter length scale physics than the staggered magnetization due to the definition of the nematic order parameter in terms of higher derivative operators. We will see below that the lack of renormalizability is a consequence of the quadratic action being incomplete; the ultraviolet divergence indicates missing operator content in the theory.

As in the previous section, we add the quartic interaction terms given in Eq. \eqref{eq:intLagrangian} to obtain the partition function
\begin{align} \label{eq:quantum_partition_full}
    \mathcal{Z}[H_1,H_2] &= \int \mathcal{D} \lambda \mathcal{D} \eta \mathcal{D} \psi_{ij}\, \exp\big( - (N-2) \mathcal{S}_{\mathrm{eff}} \big) ,\\
    \mathcal{S}_{\mathrm{eff}} &= \frac{1}{2} \mathrm{tr} \ln \big( -\chi_\perp \del_\tau^2 + \mathcal{K}(\del,\rho[H]) + \lambda \big) \nonumber\\
    & - \frac{1}{2} \int_{\vb{x},\tau} \Bigg[ \lambda + \frac{\eta^2}{\gamma_0} + \frac{\psi_1^2}{\gamma_1} + \frac{\psi_2^2}{\gamma_2} \nonumber \\
    & -  \sum_{\ell=1}^2   \sigma_\ell \left( -\chi_\perp \del_\tau^2 + \mathcal{K}(\del,\rho[H]) + \lambda \right) \sigma_\ell \Bigg] \nonumber
\end{align}
where $\mathcal{K}(\del,\rho[H])$ and $\rho_{ij}[H]$ were defined in Eq. \eqref{eq:classical_partition}, and as in Sec. \ref{sec:largeNv1}, we allow for $O(N)$ symmetry breaking to select two preferred components $\sigma_1$ and $\sigma_2$ of the staggered magnetization $\Vec{n} = (\Vec{\pi},\sigma_1,\sigma_2)$. Then, in the $N\to\infty$ limit, we obtain the effective potential
\begin{align}
    &\Gamma[\Psi_1,\Psi_2] =  - \frac{\lambda}{2} - \frac{(\eta - \rho)^2}{2\gamma_0}  \\
    &+\frac{\gamma_1 \Psi_1^2}{2} + \frac{\gamma_2 \Psi_2^2}{2}  + \Psi_1 \psi_1 + \Psi_2 \psi_2  \nonumber \\
    &+ \frac{1}{2} \sum_{\ell=1}^2 \sigma_\ell \left( -\eta \nabla^2 - \psi_{ij} \del_i \del_j + b_{ij} \del_i^2 \del_j^2 + \lambda \right) \sigma_\ell \nonumber \\
    &+\frac{g}{4} \int_{\vb{q},\omega} \ln \left( [\omega^2 + \eta \vb{q}^2 + b_{ij} q_i^2 q_j^2 +  \lambda]^2 - [\psi_{ij}q_i q_j]^2 \right)  , \nonumber
\end{align}
where $g = \chi_\perp^{-1/2}$, as before, and we have re-scaled the imaginary frequency $\omega$ by a factor of $g$. Observe that $g$ plays fundamentally the same role as temperature in the classical theory discussed above; it controls the strength of quantum fluctuations while $T$ controls classical thermal fluctuations. The details of evaluating the potential and its associated saddle-point equations are given in Appendix  \ref{app:saddles}. We simply note that the solution requires the renormalization of several parameters, $g_R$, $\rho_R$, $\gamma_{i,R}$, $\Psi_{i,R}$ and $\sigma_{i,R}$. 

\begin{figure}
    \centering
    \includegraphics[width=\linewidth]{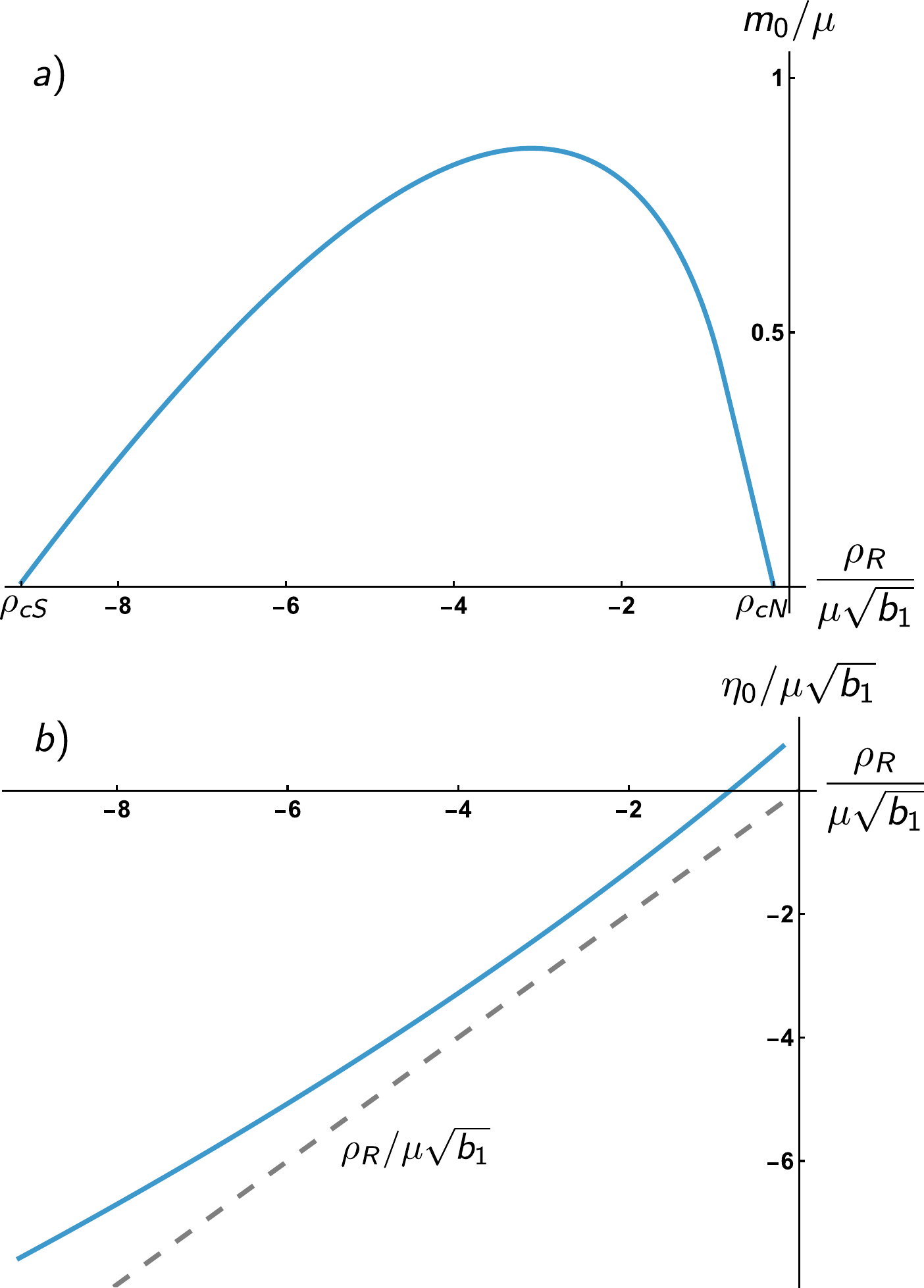}
    \caption{(a) The dimensionless mass gap $m_0/\mu$, where $m_0^2=\lambda_0$ for $\eta_0 > 0$ and $m_0^2 = \lambda_0 - \eta_0^2/4b_{\vb{Q}}$ for $\eta_0 < 0$. The N\'eel and spiral ordering transitions occur where the gap closes at $\rho_{cN}$ and $\rho_{cS}$, respectively. (b) The dimensionless physical stiffness $\eta_0/\mu\sqrt{b_1}$, both as a function of the dimensionless renormalized stiffness parameter $\rho_R/\mu\sqrt{b_1}$, for $b_2=0$, $g_R/\sqrt{b_1} = 8\pi$ and $\gamma_0/b_1=2$. In (b), the dimensionless physical stiffness is compared with the dimensionless renormalized stiffness parameter $\rho_R/\mu\sqrt{b_1}$ shown by the dashed gray line.}
    \label{fig:quantum_saddle_sols}
\end{figure}

Finally, as in the previous section, we can read off the dynamic susceptibility as a function of real frequency directly from the effective action
\begin{equation}
    \chi_n(\vb{q},\omega) = \frac{g_R}{(\omega+i0^+)^2 - (\eta\delta_{ij} + \psi_{ij})q_i q_j - b_{ij} q_i^2 q_j^2 -  \lambda},
\end{equation}
as well as the static structure factor
\begin{equation} \label{eq:quantumSq}
    S(\vb{q}) = \frac{g_R}{2\sqrt{(\eta \delta_{ij} + \psi_{ij}) q_i q_j + b_{ij} q_i^2 q_j^2 + \lambda}}.
\end{equation}
As noted above, and as shown in Fig. \ref{fig:quantumSq}, when the effective stiffness tensor $\eta \delta_{ij} + \psi_{ij}$ is positive definite, the structure factor has a single peak at $\vb{q} = 0$ which is either broad in the quantum disordered phase or a magnetic Bragg peak in the N\'eel phase. When $\eta < 0$ but $\psi_1 = \psi_2 = 0$, the structure factor has four peaks of equal weight at the incommensurate spiral wavevectors $\pm \vb{Q}_\pm$. Finally, when $\psi_1$ or $\psi_2 \neq 0$, the degeneracy between the four peaks is split, and then in the ordered spiral phase two of the peaks become magnetic Bragg peaks at the ordering wavevector.

Having obtained the effective potential, we can now follow the method of Sec. \ref{sec:classical} and use it to determine the phase diagram of the theory. We will focus on the vicinity of the spiral to quantum disordered phase transition discussed in Sec. \ref{sec:largeNv1} and then comment on the predictions the theory makes in the vicinity of the N\'eel phase.

\subsection{Quantum Disordered to Nematic Phase Transition \label{sec:quantumnematic}}

In the regime with no long-range magnetic order, $\sigma_i = 0$ and the analysis of the effective potential can be carried out in much the same way as in Sec. \ref{sec:classical}. Once again restricting our attention to the case $b_1 > b_2 \geq 0$ where it is safe to assume that $\Psi_{1,R} =0 $ and $\Psi_{2,R} \equiv \Psi$ can be non-zero, we solve the saddle-point equations order by order in a power series expansion in $\Psi$,
\begin{subequations}
\begin{align}
    \lambda(\Psi) &= \lambda(0) + \frac{1}{2!}\lambda^{(2)}(0)\Psi^2  + \dots, \\
    \eta(\Psi) &= \eta(0) + \frac{1}{2!}\eta^{(2)}(0)\Psi^2 + \dots, \\
    \psi(\Psi) &= \psi^{(1)}(0) \Psi + \frac{1}{3!}\psi^{(3)}(0) \Psi^3 +  \dots,
\end{align}
\end{subequations}
to obtain the Ginzburg-Landau expansion of the effective potential
\begin{equation}
    \Gamma[\Psi] = \Gamma[0] + \frac{\gamma_2}{2}\Psi^2 + \sum_{\ell=1}^\infty \frac{\psi^{(2\ell-1)}(0)}{(2\ell)!} \Psi^{2\ell}.
\end{equation}

Beginning at zeroth order, we solve the saddle-point equations numerically exactly for $\lambda_0 = \lambda(0)$ and $\eta_0 = \eta(0)$ as functions of $\rho_R$, with the results shown in Fig. \ref{fig:quantum_saddle_sols}. Similarly to the correlation length in Sec. \ref{sec:classical}, the mass gap is measured relative to the minimum of the dispersion: For $\eta_0 > 0$, the gap is $m=\lambda_0^{1/2}$ and the short-range antiferromagnetic correlations will be commensurate and N\'eel-like. For $\eta_0 < 0$, the gap is $m = (\lambda_0 - \eta_0^2/4b_{\vb{Q}})^{1/2}$ as the dispersion develops minima along the preferred spiral directions $\vb{Q}_\pm$, signaling short-range incommensurate correlations. The mass gap shown in Fig. \ref{fig:quantum_saddle_sols} (a) vanishes linearly at the quantum disordered to N\'eel ($\rho_{cN}/\mu\sqrt{b_1} \approx -0.19$) and spiral to quantum disordered ($\rho_{cS}/\mu\sqrt{b_1} \approx -9.15$) transitions with different gradients at each point, in agreement with the analysis of Sec. \ref{sec:largeNv1} which showed that the behavior at the spiral to quantum disordered transition depends strongly on the anisotropy ratio $b_2/b_1$. Notably, as seen in Fig. \ref{fig:quantum_saddle_sols} (b), the physical local stiffness $\eta_0$ is renormalized upwards from the parameter $\rho_R$, extending the N\'eel phase into the region of $\rho_R < 0$ and yielding the $\rho_{cN} < 0$ visible in the left panel.

\begin{figure}
    \centering
    \includegraphics[width=\linewidth]{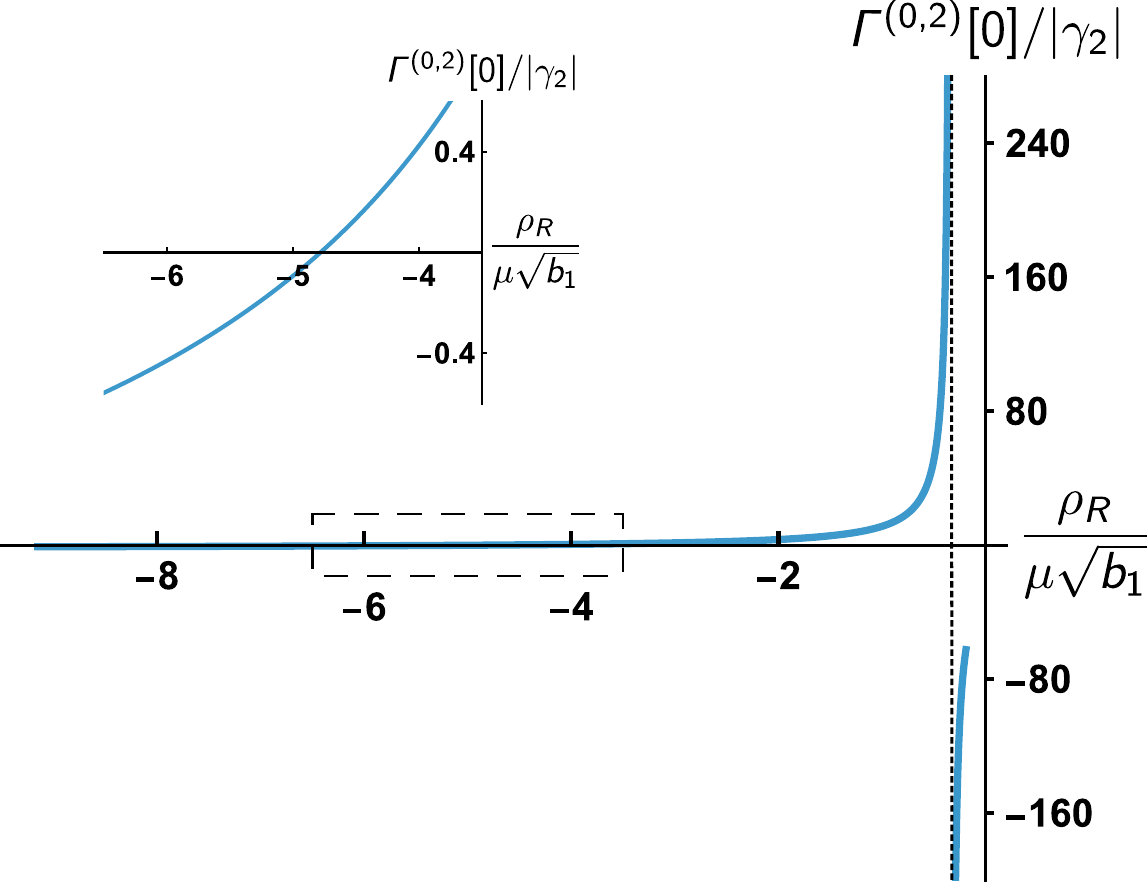}
    \caption{Second order coefficient in the Ginzburg-Landau expansion of the effective potential for $b_2=0$, $g_R/\sqrt{b_1} = 8\pi$, $\gamma_0/b_1=2$ and $\gamma_2/\gamma_0 = -1/8$. The inset shows a magnified view of the region in the dashed box. The dotted vertical line indicates the location of the pole in $\Gamma^{(0,2)}[0]$.}
    \label{fig:quantum_Gamma2}
\end{figure}

To next order in the Ginzburg-Landau expansion, we find the second-order coefficient
\begin{align} \label{eq:quantum_gamma2}
    &\Gamma^{(0,2)}[0] = \gamma_2 + \frac{1}{g_R} \Bigg( \int_0^{2\pi} \frac{d \theta}{2\pi} \frac{\sin^2(2\theta)}{32\pi b_\theta^{3/2}} \\
    &\times\left[ \ln\left(\frac{\sqrt{4b_\theta\mu^2}}{\sqrt{4b_\theta\lambda_0} + \eta_0 }\right) - \frac{\eta_0}{\sqrt{4b_\theta \lambda_0} + \eta_0 }  \right]  \Bigg)^{-1}, \nonumber
\end{align}
which is shown in Fig. \ref{fig:quantum_Gamma2} for $\gamma_2/\gamma_0=-1/8$. We note four important features:
\begin{enumerate}[label=(\roman*),leftmargin=0pt,itemindent=25pt]
    \item $\Gamma^{(0,2)}[0]$ becomes negative smoothly at $\rho_{c\Psi}/\mu\sqrt{b_1} \approx -4.8$, suggesting a continuous nematic phase transition from the quantum disordered phase into an Ising nematic phase before any long-range spiral order sets in.
    \item We see from Fig. \ref{fig:quantum_saddle_sols} (b) that $\eta_0$ becomes negative at $\rho_R/\mu\sqrt{b_1}\approx -0.8$, and hence, short-range incommensurate correlations set in well before the nematic phase transition.
    \item $\Gamma^{(0,2)}[0] = \gamma_2$ at the original quantum disordered to spiral transition $\rho_{cS}$. Therefore, when $\gamma_2 = 0$ (that is, in the absence of the quartic interactions), the nematic susceptibility can only diverge at the onset of spiral order. Hence, any negative value of $\gamma_2$ is sufficient to stabilize nematic order as one is correspondingly close to the spiral transition. This mirrors the behavior of the classical system discussed in Section \ref{sec:classical}.
    \item $\Gamma^{(0,2)}[0]$ also has a singularity at $\rho_{R}/\mu\sqrt{b_1} \approx -0.3$, becoming negative before the original ``unperturbed'' N\'eel transition. This is suggestive of a vanishing nematic correlation length $\xi_\Psi \to 0$, as might occur at a strongly first-order phase transition preempting the conventional quantum disordered to N\'eel transition.
\end{enumerate}
For the rest of this section, we will focus on the behavior close to the unperturbed spiral phase, and comment on the unusual behavior near the N\'eel phase in Section \ref{sec:neel}.

\begin{figure}
    \centering
    \includegraphics[width=\linewidth]{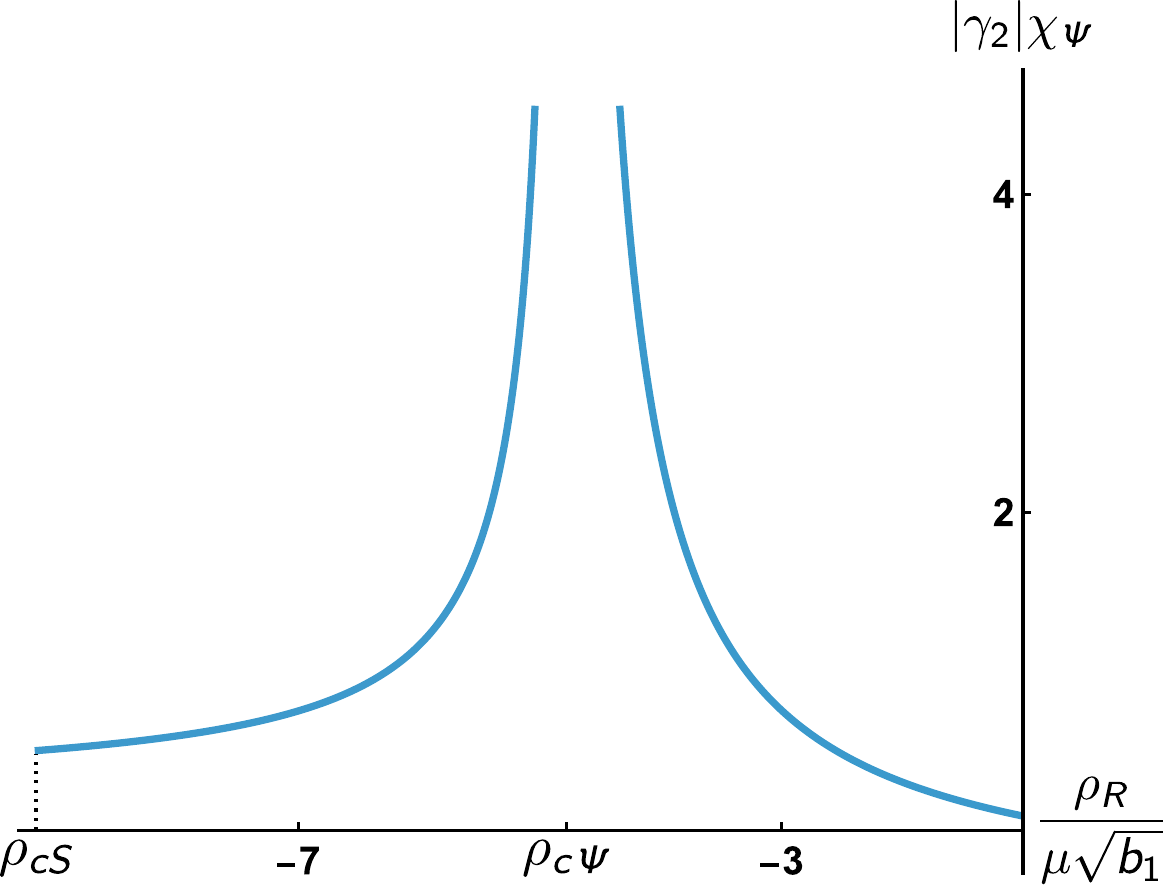}
    \caption{Quantum nematic susceptibility as a function of $\rho_R/\mu\sqrt{b_1}$ for $b_2=0$, $g_R/\sqrt{b_1} = 8\pi$, $\gamma_0/b_1=2$ and $\gamma_2/\gamma_0 = -1/8$. Here $\rho_{c\Psi}/\mu\sqrt{b_1} \approx -4.78$. The location of the spiral ordering transition when $\Psi=0$ is marked as $\rho_{cS}$; generally $\rho_{cS}$ will depend on $\Psi$, as discussed in Sec. \ref{sec:spiral}. For $\rho_R < \rho_{c\Psi}$ the susceptibility in the nematic phase follows from the Ginzburg-Landau analysis as $\chi_\Psi = \vert2\Gamma^{(0,2)}[0]\vert^{-1}$. }
    \label{fig:quantum_suscept}
\end{figure}

By solving the linearized saddle-point equations numerically up to $\mathcal{O}(\Psi^3)$, we also obtain $\Gamma^{(0,4)}[0] = \psi^{(3)}(0)$ numerically. Unlike in the classical case discussed in the previous section, here $\Gamma^{(0,4)}[0]$ is strictly positive throughout the entire quantum disordered phase, though it does diverge near the N\'eel phase at the same point as $\Gamma^{(0,2)}[0]$. From this, we deduce that there is indeed a continuous phase transition from the quantum disordered phase to the nematic, and we plot the nematic susceptibility in the vicinity of the transition in Fig. \ref{fig:quantum_suscept}. This is our second main result and the central message of this work; by including all interactions allowed by symmetry, we have shown that a quantum Ising nematic phase can be stabilized. Additionally, it is clear that the nematic phase is a vestige of the spiral order which is melted by strong infrared quantum fluctuations: As $\gamma_2$ is tuned from being positive to negative, the nematic phase emerges directly from the adjoining spiral phase. Therefore, this model provides a clear example of a manifestly quantum, zero temperature vestigial order.

\begin{figure}[tb]
    \centering
    \includegraphics[width=\linewidth]{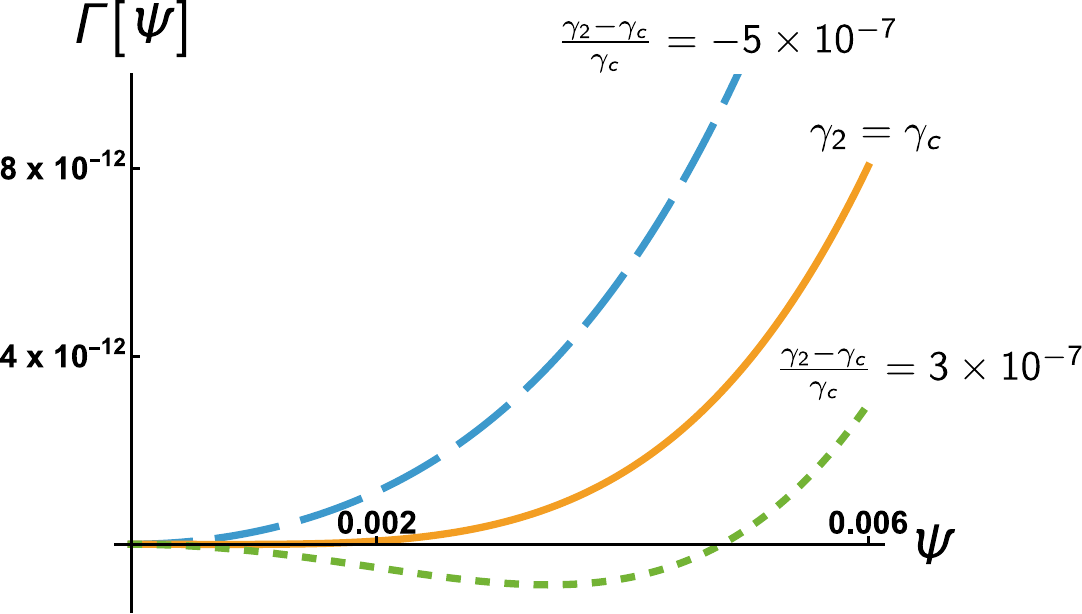}
    \caption{The exact effective potential $\Gamma[\Psi]$ as a function of the nematic order parameter $\Psi$ (both in arbitrary units), given by Eq. \eqref{eq:quantum_eff_pot_nematic}, for $b_2=0$, $g_R/\sqrt{b_1} = 8\pi$, $\gamma_0/b_1=2$, and $\rho_R/\mu\sqrt{b_1} = -4.8$.}
    \label{fig:pot_exact}
\end{figure}

Finally, while the Ginzburg-Landau expansion provides useful insight into the nature of the phase diagram, it is only a power series expansion of the effective potential. We can also solve the exact nonlinear saddle-point equations to obtain $\psi(\Psi)$. Then, by using Eq. \eqref{eq:legendre_identity}, which we can integrate to write as
\begin{equation} \label{eq:quantum_eff_pot_nematic}
    \Gamma[\Psi] = \frac{\gamma_2}{2}\Psi^2 + \int_0^\Psi d X \psi(X),
\end{equation}
the full effective potential follows from $\psi(\Psi)$ by a simple numerical integral (as a Riemann sum). The result is shown in Fig. \ref{fig:pot_exact} for $\rho_R/\mu\sqrt{b_1} = -4.8$ for three values of $\gamma_2$ in the vicinity of the nematic phase transition.

\subsection{Nematic to Spiral Phase Transition \label{sec:spiral}}

Having established the existence of an Ising nematic phase within the quantum disordered regime, we will now show how the nematic order influences the original quantum disordered to spiral transition discussed in Sec. \ref{sec:largeNv1}.

Throughout this work so far we have treated the components $\sigma_\ell$ of the field $\vec{n} = (\vec{\pi},\sigma_1,\sigma_2)$ which were not integrated out in the functional integral simply as variational parameters, much like the Hubbard-Stratonovich fields $\lambda$, $\eta$, and $\psi_i$. However, unlike the Hubbard-Stratonovich fields, as components of the staggered magnetization they are physical observables. Therefore, one can view the large-$N$ limit of the free energy as an effective potential for the magnetic order parameters:
\begin{align}
    &\Gamma[\sigma_\ell] = -\frac{\psi_1^2}{2\gamma_1} - \frac{\psi_2^2}{2\gamma_2}  - \frac{\lambda}{2} - \frac{(\eta - \rho)^2}{2\gamma_0} \\
    &+ \frac{1}{2} \sum_{i=\ell}^2 \sigma_\ell \left( -\eta \nabla^2 - \psi_{ij} \del_i \del_j + b_{ij} \del_i^2 \del_j^2 + \lambda \right) \sigma_\ell \nonumber \\
    &+\frac{g}{4} \int_{\vb{q},\omega} \ln \left( [\omega^2 + \eta \vb{q}^2 + b_{ij} q_i^2 q_j^2 +  \lambda]^2 - [\psi_{ij}q_i q_j]^2 \right), \nonumber
\end{align}
where now $\lambda = \lambda(\sigma_\ell)$, $\eta = \eta(\sigma_\ell)$, and $\psi_i = \psi_i(\sigma_\ell)$. In the presence of nematic ordering with $\psi_1 = 0$ and $\psi_2 \equiv \psi$, the degeneracy between the original two spiral directions given in Eq. \eqref{eq:spiralQ} is broken and there is only one minimal spiral solution $\sigma_1 = \sigma \cos(\vb{Q}_\psi \cdot \vb{x})$ and $\sigma_2 = \sigma \sin(\vb{Q}_\psi \cdot \vb{x})$, where the ordering wavevector is
\begin{equation}
    \vb{Q}_\psi = \sqrt{\frac{\abs{\eta - \abs{\psi}}}{2 b_{\vb{Q}}}} \big( \hat{\vb{x}} - \mathrm{sign}(\psi) \hat{\vb{y}} \big) .
\end{equation}
One can then show using the saddle-point equations that
\begin{equation}\label{eq:quantum_eff_pot_derivative}
    \frac{\delta \Gamma[\sigma]}{\delta \sigma} = \left(  \lambda(\sigma) - \frac{(\eta(\sigma) - \abs{\psi(\sigma)})^2}{4 b_{\vb{Q}}} \right) \sigma .
\end{equation}
It follows that the effective potential as a function of $\sigma$ can be determined directly from the solutions $\lambda(\sigma)$, $\eta(\sigma)$ and $\psi(\sigma)$ to the saddle-point equations,
\begin{equation} \label{eq:quantum_eff_pot_spiral}
    \Gamma[\sigma] = \int_0^\sigma d Y \left(  \lambda(Y) - \frac{(\eta(Y) - \abs{\psi(Y)})^2}{4 b_{\vb{Q}}} \right)  .
\end{equation}

Before proceeding, we note some general properties of the potential. First, just as in Sec. \ref{sec:quantumnematic}, positive-definiteness of the action guarantees that the gap must always satisfy
\begin{equation} \label{eq:gap_inequality}
    \lambda \geq \frac{(\eta - \abs{\psi})^2}{4b_{\vb{Q}}},
\end{equation}
which suggests that $\Gamma[\sigma]$ must be a non-decreasing function of $\sigma$; the minimum value the potential can attain is $0$. Taken at face value, this would have the strange implication that in the spiral phase, where a finite condensate $\sigma_0 > 0$ exists, the potential $\Gamma[\sigma]$ is flat in the range $0 \leq \sigma \leq \sigma_0$. However, the derivation above assumes that the solutions to the saddle-point equations $\lambda(\sigma)$, $\eta(\sigma)$ and $\psi(\sigma)$ exist for all values of $\sigma$, which is not necessarily the case. To see this, we look for solutions where $\Gamma[\sigma] = 0$ identically by fixing $\lambda = (\eta - \abs{\psi})^2/4b_{\vb{Q}}$. However, this eliminates a variable from the system of saddle-point equations, rendering them over-specified and fixing the unique value $\sigma = \sigma_0$ where a solution exists. Therefore, in the spiral phase $\Gamma[\sigma]$ is undefined in the range $0 \leq \sigma < \sigma_0$. This is not unusual and is simply a consequence of working at fixed order parameter instead of fixed external field. By analogy to the thermodynamics of a gas held either at constant pressure or at constant volume, we deduce that the range $0 \leq \sigma < \sigma_0$ corresponds to a thermodynamically unstable mixed-phase regime. By setting $\Gamma[0] = 0$ by hand in this whole regime we are essentially performing a Maxwell construction of the physical effective potential.\footnote{Recall that in classical thermodynamics the Maxwell construction sacrifices analyticity of the equation of state to recover a physical single-valued function.}

\begin{figure}[tb]
    \centering
    \includegraphics[width=\linewidth]{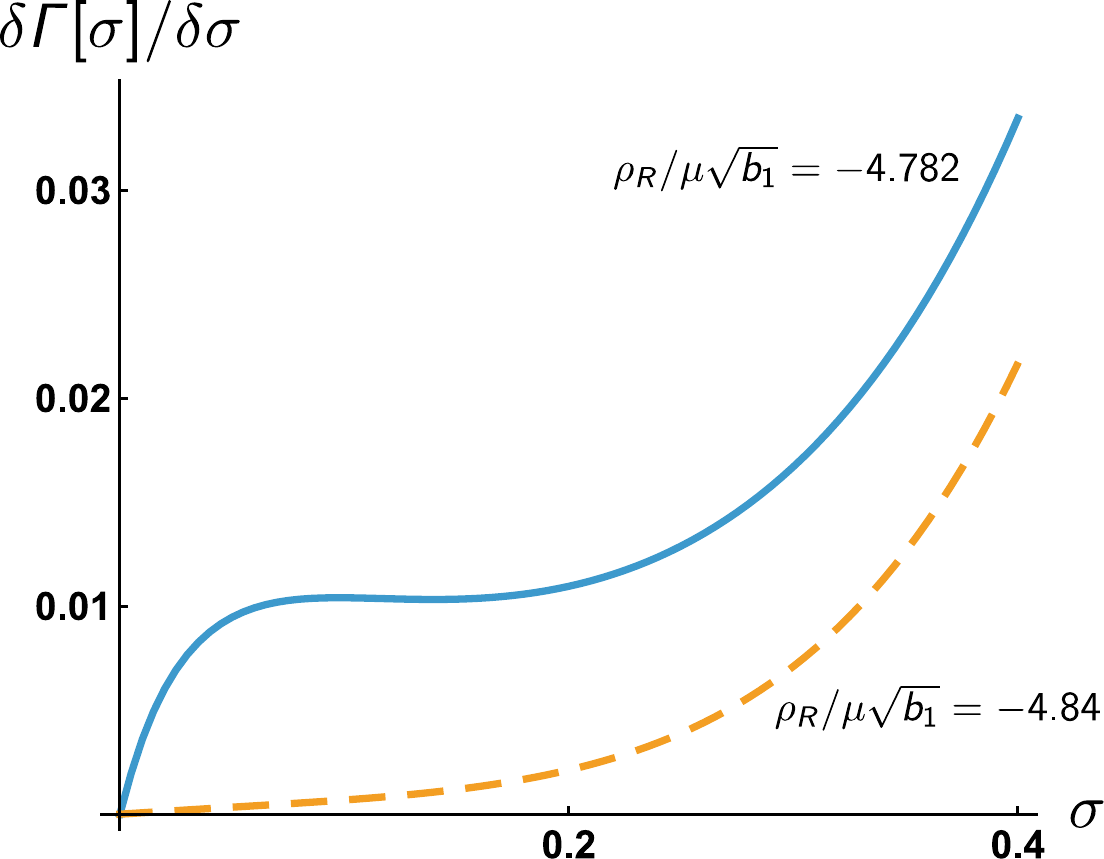}
    \caption{The first derivative of the effective potential $\Gamma[\sigma]$ as a function of the spiral amplitude $\sigma$, given by Eq. \eqref{eq:quantum_eff_pot_derivative}, evaluated within the nematic phase with parameters $b_2=0$, $g_R/\sqrt{b_1} = 8\pi$, $\gamma_0/b_1=2$ and $\gamma_2/\gamma_0 = -1/8$. The spiral transition is at $\rho_{cS}/\mu\sqrt{b_1} \approx 4.855$.}
    \label{fig:magpot}
\end{figure}

\begin{figure}[tb]
    \centering
    \includegraphics[width=\linewidth]{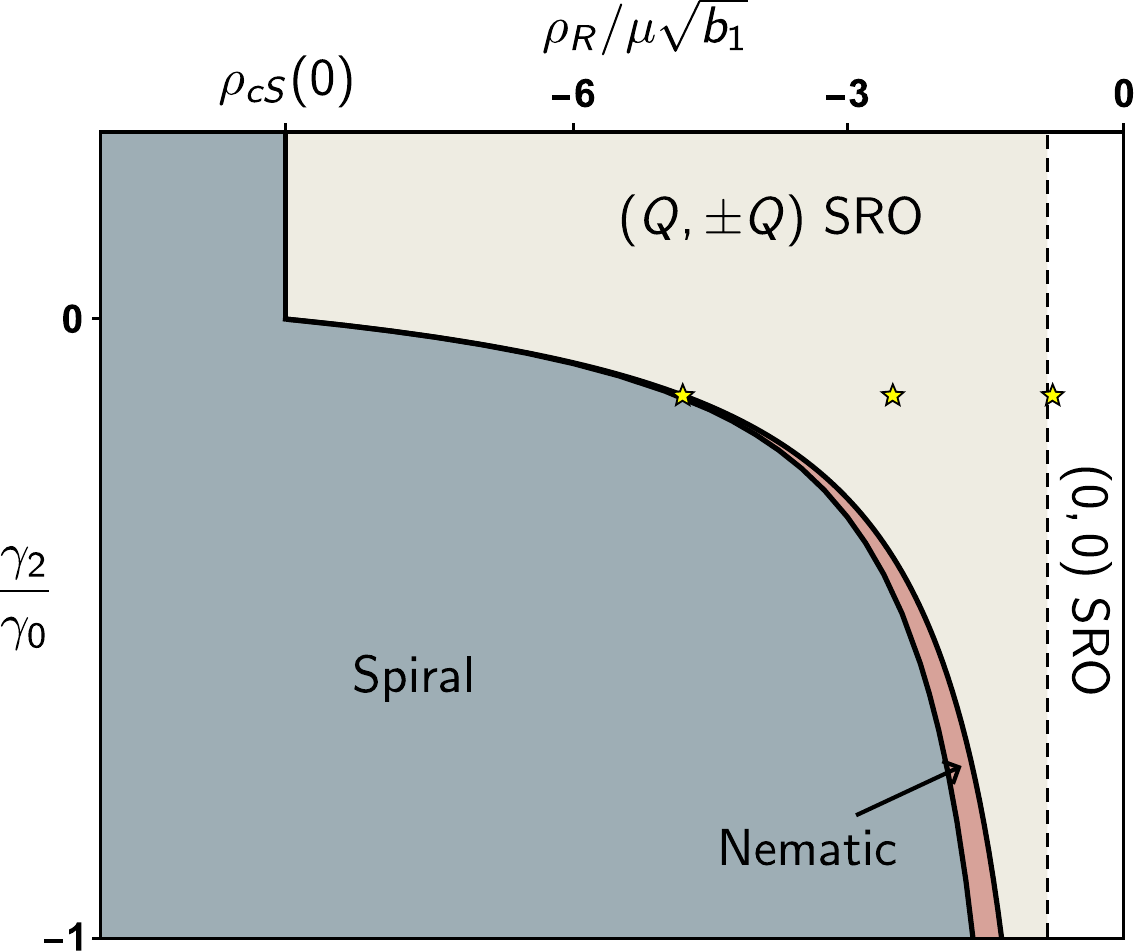}
    \caption{$N=\infty$ phase diagram of the theory in Eq. \eqref{eq:quantum_partition_full} as a function of the dimensionless stiffness parameter $\rho_R/\mu\sqrt{b_1}$ and the interaction strength $\gamma_2/\gamma_0$. The parameters used are $b_2=0$, $g_R/\sqrt{b_1} = 8\pi$, and $\gamma_0/b_1=2$. The blue shaded region corresponds to the long-range ordered spiral phase, the red region to the vestigial Ising-nematic phase, the beige region to the $C_{4v}$ symmetric quantum disordered regime with incommensurate short-range spiral fluctuations, and the white region to the $C_{4v}$ symmetric quantum disordered regime with commensurate fluctuations. The vertical dashed line denotes the crossover between these last two regimes and occurs when $\eta_0(\rho_R) = 0$ (see Fig. \ref{fig:quantum_saddle_sols} (b)). The N\'eel phase (not shown) lies to the right of the commensurate quantum disordered regime. The three yellow stars denote the points at which the static structure factor is evaluated in Fig. \ref{fig:quantumSq}.}
    \label{fig:phaseDiagram}
\end{figure}

\begin{figure*}
    \centering
    \includegraphics[width=\linewidth]{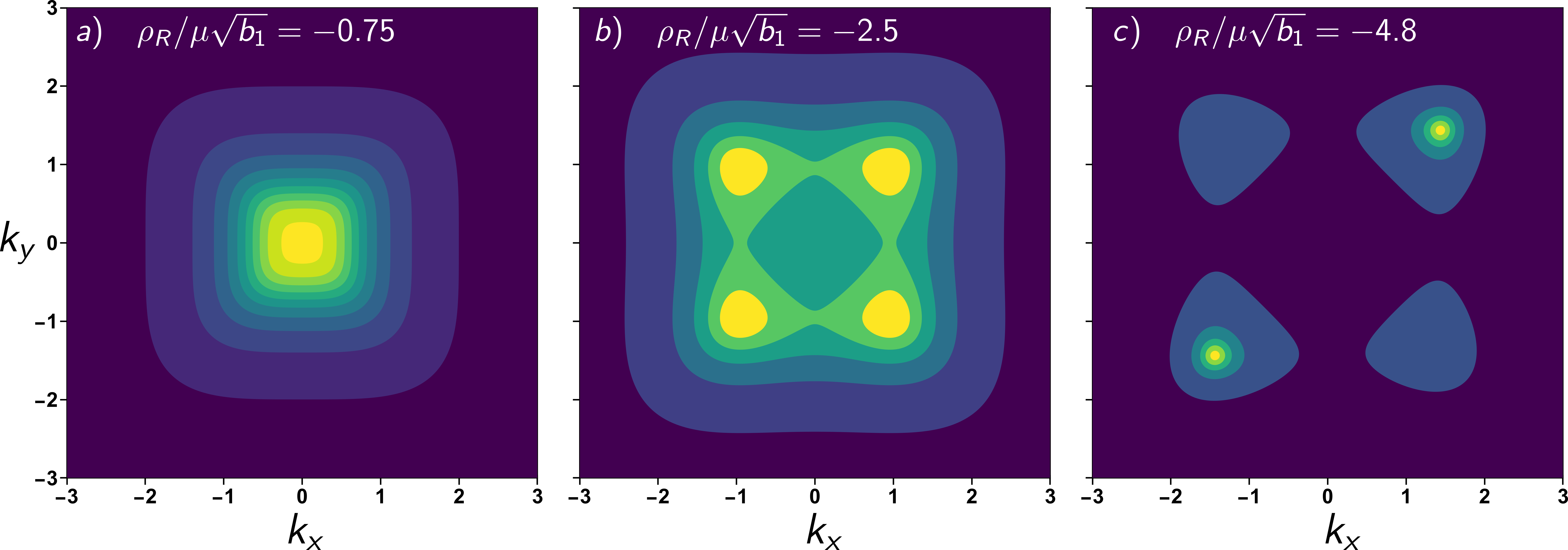}
    \caption{Static structure factor $S(\vb{q})$ given by Eq. \eqref{eq:quantumSq}, where $\vb{k} = (\sqrt{b_1}/\mu)\vb{q}$ is the dimensionless momentum measured from $(\pi,\pi$), and $b_2=0$, $g_R/\sqrt{b_1} = 8\pi$, $\gamma_0/b_1=2$ and $\gamma_2/\gamma_0 = -1/8$. The three sub-figures correspond to the points in the phase diagram in Fig. \ref{fig:phaseDiagram} marked with yellow stars. Panel (a) is in the regime of the quantum disordered phase with commensurate fluctuations, panel (b) is in the regime with incommensurate fluctuations, and panel (c) is in the nematic phase in one of the two degenerate ground states (the other being related by a $\pi/4$ rotation).}
    \label{fig:quantumSq}
\end{figure*}

In Fig. \ref{fig:magpot} we plot $\delta\Gamma[\sigma]/\delta\sigma$ for two different parameter values approaching the spiral ordering transition from the nematic side. For smaller values of $\gamma_2$, closer to the nematic to quantum disordered transition, the derivative of the potential has a complicated functional dependence on $\sigma$, even displaying a weak local minimum at finite $\sigma$. However, on decreasing $\gamma_2$ further toward the spiral transition, the derivative becomes a strictly increasing function of $\sigma$. Therefore, close enough to the transition, $\delta^2\Gamma[\sigma]/\delta\sigma^2$ can only vanish at $\sigma = 0$, and hence, we infer that the spiral transition must be continuous. We also observe that the nematicity drastically shifts the location of the spiral transition from the unperturbed value $\rho_{cS}/\mu\sqrt{b_1} \approx -9.15$ to, in the the case of $\gamma_2/\gamma_0=-1/8$, $\rho_{cS}(\gamma_2)/\mu\sqrt{b_1} \approx -4.855$. One can repeat this analysis for $\gamma_2 > 0$ to find that $\delta\Gamma[\sigma]/\delta\sigma$ is always a strictly increasing function of $\sigma$ in this regime. As a caveat, we note that we have not proven definitively that the nematic to spiral transition is continuous for all possible ranges of parameters, rather we have found no evidence of a first order transition in the physical range $0 < \abs{\gamma_2} < \gamma_0$.

Having confirmed that the spiral transition is continuous, it is straightforward to solve the saddle-point equations for the critical stiffness $\rho_{cS}(\gamma_2)$. All of the results of this section can then be summarized in the phase diagram in Fig. \ref{fig:phaseDiagram}, which shows the evolution of the spiral, nematic and quantum disordered phases as a function of stiffness parameter $\rho_R$ and nematic interaction strength $\gamma_2$; recall that the local stiffness is related to the frustration $J_3/J_1$. For $\gamma_2 > 0$, there is no nematic phase, and the system undergoes a continuous phase transition from a quantum disordered phase with short-ranged incommensurate correlations to a long-range ordered spiral phase. For $\gamma_2 < 0$, there is a very narrow vestigial Ising nematic phase between the spiral phase and the quantum disordered regime with short-range incommensurate correlations; the higher-order quartic interactions stabilize long-range $\mathbb{Z}_2$ order which persists despite the destruction of magnetic order. The lower bound of the phase diagram is at $\gamma_2/\gamma_0 = -1$ since, as noted at the beginning of Sec. \ref{sec:effectivefieldtheory}, stability of the theory requires $\gamma_0 + \gamma_1 + \gamma_2 > 0$.

\subsection{Near the N\'eel Phase \label{sec:neel}}

Finally, we conclude this section with a discussion of the unusual divergence in the second-order Ginzburg-Landau coefficient $\Gamma^{(0,2)}[0]$ in the vicinity of the N\'eel phase, as shown in Fig. \ref{fig:quantum_Gamma2}. We noted above that a diverging Ginzburg-Landau coefficient was suggestive of the correlation length vanishing at a strongly first-order phase transition. In fact, by applying a similar analysis as that described in the previous subsection to calculate the effective potential in the ordered phase, we find that the system develops N\'eel order with an anisotropic stiffness. Since this behavior is unusual, particularly because it does not reflect any known phase of the $J_1$-$J_3$ Heisenberg model, it is worth addressing why we believe it to be an artifact of our method.

First, we note that it is clear from Eq. \eqref{eq:quantum_gamma2} that this divergence is independent of $\gamma_2$; it persists even in the absence of the quartic interaction which gave rise to nematicity. Similarly, it is also straightforward to find that the divergence is not a consequence of anisotropy.\footnote{The location of the divergence as a function of the stiffness parameter $\rho_R$ can be found exactly in the limit $\gamma_0 \to 0$ when $b_1 = b_2$. A careful analysis confirms that the divergence always preempts the N\'eel transition, regardless of the value of $g_R$.} The questions, then, are i) what is the origin of the divergence, and ii) what are its implications for the rest of our analysis?  

First, recall the central idea of how the low energy effective action was constructed: A spiral phase can be stabilized at a negative value of the local stiffness parameter only if higher order contributions in spatial derivatives to the elastic energy are included. However, a consistent effective field theory must include all symmetry-allowed operators with the same scaling dimensions, and so we were led to consider quartic non-elastic contributions at the same order in spatial derivatives; these contributions can quite generally be expected to be generated under renormalization as one flows from an ultraviolet lattice model to the infrared continuum description. We then showed throughout this work that these new operators were crucial for stabilizing Ising nematic order in the vicinity of the spiral transition. 

However, when the (renormalized) stiffness parameter is positive, all higher-order derivative contributions are RG irrelevant and the infrared scaling is dominated by the behavior of the conventional $(2+1)$-dimensional NLSM. Crucially, our analysis does not (and cannot) naturally account for this difference in scaling behavior across the two regimes. We have assumed that the seven parameters of the effective field theory can all be tuned independently, because it is impossible to correctly determine their actual dependence on the four parameters $J_1$, $J_2$, $J_3$ and $S$ of the lattice Heisenberg model. Additionally, the large-$N$ technique used throughout this work is an approximation that rests on the assumption that the only qualitatively important interactions are those captured by one-loop diagrams. By nonperturbatively resumming an infinite number of one-loop diagrams from an irrelevant operator, the method may give too much weight to the contributions of these higher-order operators. Therefore, we conclude that there is strong evidence for the validity of our method in the vicinity of the spiral phase where all the included operators are necessary, but that physical conclusions should not be drawn from the results in proximity to the N\'eel phase.

\section{Discussion \label{sec:disc}}

In this paper, we demonstrated the existence of a vestigial Ising nematic phase, adjacent to the parent spiral phase, in the quantum $J_1$-$J_3$ Heisenberg model at zero temperature. To do this, we considered a continuum effective field theory, solvable in the large-$N$ limit, and used the discrepancy observed in Ref. \cite{Capriotti2004} between classical Monte Carlo simulations and the corresponding analytical treatment to argue that stabilizing nematic order requires going beyond the previously considered elastic approximation. By including all possible symmetry-allowed operators up to fourth-order in spatial derivatives, we were able to derive the phase diagram of the theory and the staggered magnetization structure factor nonperturbatively in the strength of the couplings. Our work complements the distinct approach taken by Read and Sachdev, in which they constructed a large-$N$ limit based on $Sp(N)$ ``spins'' in a lattice model \cite{Sachdev1991}. Each approach captures different essential physics, and together paint a more detailed theoretical understanding of 2D frustrated magnetism.

\begin{figure}[bt]
    \centering
    \includegraphics[width=\linewidth]{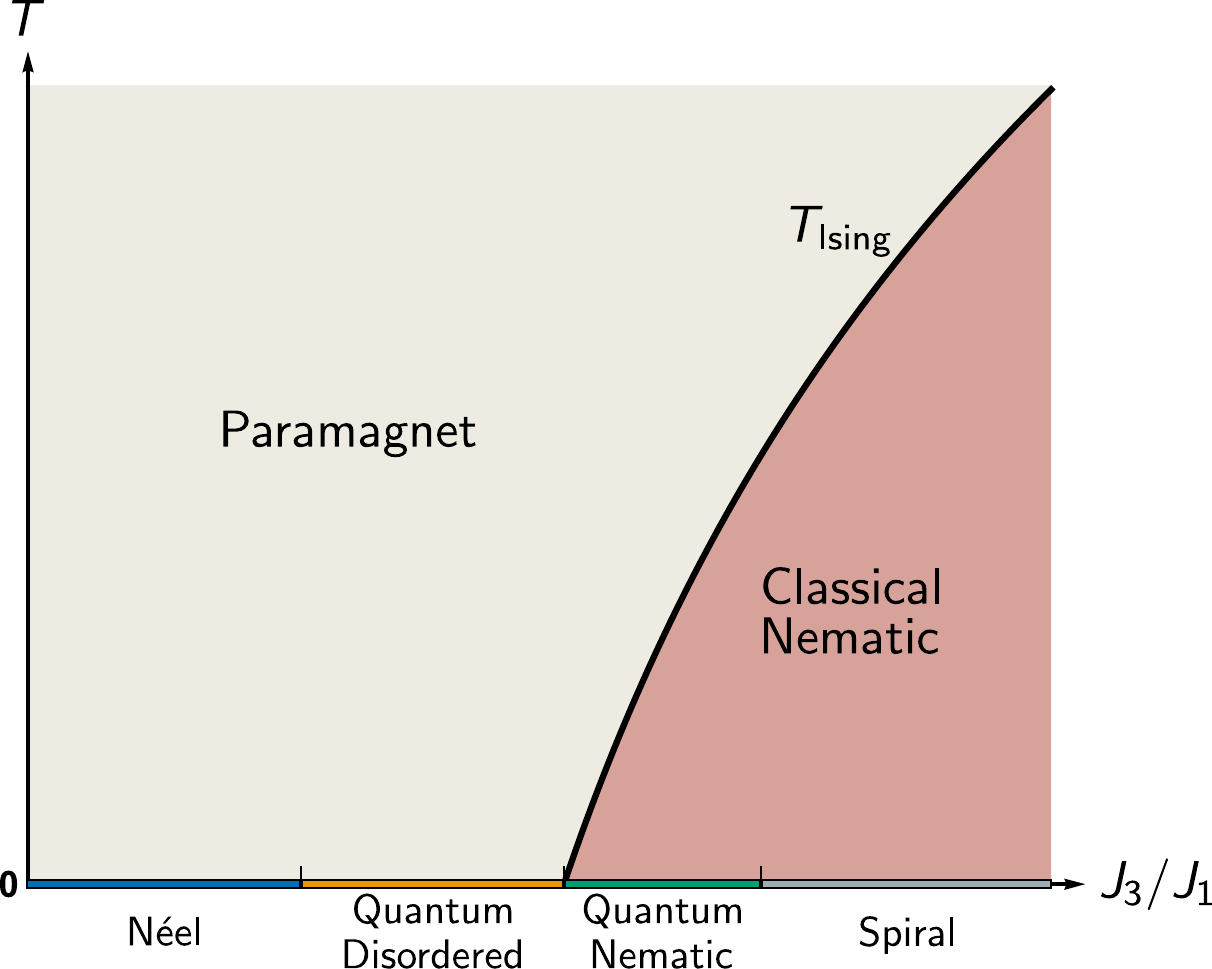}
    \caption{Schematic $S\to\infty$ phase diagram of the $J_1$-$J_3$ model summarizing the main results of this work. For $T>0$, our results in Sec. \ref{sec:classical} reproduced the Ising nematic transition first described in Ref. \cite{Capriotti2004}. We showed in Sec. \ref{sec:quantumvestigial} that an analogous quantum Ising nematic transition also occurs at $T=0$.}
    \label{fig:summary}
\end{figure}

Our first main result demonstrated that effective field theory and the large-$N$ limit could be used to describe the Ising nematic phase of the classical $J_1$-$J_3$ model. We examined this simpler scenario in great detail for two reasons: First, it had previously been reported that the $N=\infty$ nematic susceptibility could only diverge at $T=0$ \cite{Capriotti2004}. Second, with known numerical results to compare with, we were able to confirm the validity of our technique, proving that a stable nematic phase requires higher-order interactions. Next, we derived the $N=\infty$ phase diagram of the quantum $J_1$-$J_3$ model as a function of the spin stiffness parameter and the strength of the nematicity-inducing interaction. This demonstrated the existence of a zero temperature Ising nematic phase, reflecting the survival of discrete rotation symmetry breaking even after long-range magnetic order is destroyed by strong infrared quantum fluctuations in the vicinity of the classical LP. By illustrating that the nematic phase is generally a sub-region of the quantum disordered regime with incommensurate fluctuations, we also showed that nematic order is related to but not identical to the existence of short-range spiral correlations. In addition, we found that the quantum nematic phase is the natural continuation to zero temperature of the classical nematic known from Monte Carlo studies, since both exist for the same range of field theory parameters. In Fig. \ref{fig:summary} we summarize these results schematically as a phase diagram for the $J_1$-$J_3$ model in the $S\to\infty$ limit as a function of temperature $T$ and frustration $J_3$. Although a calculation of the low-temperature quantum correction to the nematic critical temperature $T_{\mathrm{Ising}}$ was beyond the scope of this work, the discrete nature of the broken rotational symmetry implies gapped excitations, and hence rules out a discontinuity in the phase diagram at $T=0$. Therefore, we can deduce that $T_{\mathrm{Ising}}(J_3)$ must terminate at the transition between the conventional quantum disordered and quantum nematic phases.

Authors of almost all prior works on both the classical and quantum $J_1$-$J_3$ models have considered only their lattice realizations and sought to understand the microscopic details of their symmetry-breaking phases. However, we have shown that the structure of the phase diagram and the nature of magnetic correlations in the $J_1$-$J_3$ model are predominantly determined by infrared-scale physics. For example, we found qualitative consistency with the finite PEPS calculations for the $S=1/2$ model reported in Ref. \cite{Liu2024}. In that work, the authors made numerical measurements of local order parameters, finding the following sequence of phases on increasing $J_3$: (i) N\'eel, (ii) QSL (iii) $C_{4v}$ symmetric VBS, (iv) VBS with broken discrete rotation symmetry, and (v) spiral; see Fig. 2 of that work. While our effective field theory cannot resolve the microscopic nature of phases with no long-range magnetic order (e.g., whether they break lattice translation symmetries), we predict an identical sequence of broken magnetic and discrete rotation symmetries and our quantum vestigial nematic phase is morally similar to the ``mixed columnar-plaquette VBS'' phase with broken $C_{4v}$ symmetry (though the authors of Ref. \cite{Liu2024} reported a first order nematic to spiral transition). Further, the evolution of the static spin structure factor with increasing $J_3$ shown in Fig. 5 of that work mirrors the behavior we obtained for the staggered magnetization response (shown in Fig. \ref{fig:quantumSq}). 

We note that there have been many experimental measurements of nematic phases in a variety of materials (see Refs. \cite{Fradkin2010,Fernandes-2019} for reviews). This includes the electronic liquid crystal state measured in the underdoped cuprate YBa$_2$Cu$_3$O$_{6.45}$ using neutron scattering \cite{Hinkov2008}, where the nematic is also proximate to a low-temperature stripe phase (that has never been observed). In that case, the onset of nematicity was measured via the asymmetric splitting of the peak at $(\pi,\pi)$ in the dynamic structure factor, in large part due to an underlying orthorhombicity. However, in our case, the correct experimental measure would be the difference in intensity along the two preferred axes $\Psi \sim S(\vb{Q}_+) - S(\vb{Q}_-)$.

Finally, in this paper, we have considered only the square-lattice $J_1$-$J_3$ model. However, our approach is general enough to be applicable to any system with a phase transition driven by infrared scale physics (such as the softening of Goldstone modes).

\begin{acknowledgments}
The authors thank L. Radzihovsky, O. Sushkov and S. Sachdev for useful discussions. This work was supported in part by the US National Science Foundation through Grant No. DMR-2225920 at the University of Illinois.
\end{acknowledgments}

\renewcommand{\theequation}{\Alph{section}.\arabic{equation}}

\appendix

\section{Linear Spin-Wave Spectrum of \texorpdfstring{$J_1$-$J_2$-$J_3$}{J1-J2-J3} Model\label{app:spinwave}}

In this appendix, we review a derivation of the semiclassical linear spin-wave spectrum of the $J_1$-$J_2$-$J_3$ model and find a mapping between the parameters of the lattice model and continuum effective field theory. 

Consider the Hamiltonian for the $J_1$-$J_2$-$J_3$ model on a square lattice (with spacing $a$),
\begin{equation}
    H = J_1 \sum_{\langle i,j\rangle_1} \vb{S}_i \cdot \vb{S}_j + J_2 \sum_{\langle i,j\rangle_2} \vb{S}_i \cdot \vb{S}_j + J_3 \sum_{\langle i,j\rangle_3} \vb{S}_i \cdot \vb{S}_j ,
\end{equation}
where $\langle i, j \rangle_\alpha$ denotes summation over the $\alpha$\textsuperscript{th} nearest neighbors. We expand about a classical N\'eel state with $S^{(z)} = \pm  S$ on $A$ and $B$ sublattices of the square lattice, respectively. After a spatial Fourier transform, the linearized Heisenberg equations of motion can be written as a matrix equation
\begin{equation} \label{eq:heisenbergEq}
    \frac{d S^{(\alpha)}_{\vb{k},\mu}}{d t}  = 4iS [ \Tilde{J}(\vb{k}) \sigma^{(z)} + i J_1(\vb{k}) \sigma^{(y)}]_{\alpha\beta} \tau^{(y)}_{\mu\nu} S^{(\beta)}_{\vb{k},\nu}
\end{equation}
where $\alpha$, $\beta$ run over the $x$ and $y$ components of spin and $\mu$, $\nu$ run over the $A$ and $B$ sublattices, $\sigma$ and $\tau$ are the Pauli matrices on the spin component and sublattice spaces, respectively, and we have defined 
\begin{subequations}
\begin{align}
    \Tilde{J}(\vb{k}) &= J_1 - [1-\gamma_2(\vb{k})]J_2 - [1-\gamma_3(\vb{k})] J_3,\\
    J_1 (\vb{k}) &= \gamma_1(\vb{k}) J_1 ,
\end{align}
\end{subequations}
and
\begin{equation}
    \gamma_\mu(\vb{k}) = \frac{1}{4} \sum_{\lbrace \vb*{\delta}_j \rbrace_\mu} e^{i \vb{k} \cdot \vb*{\delta}_j},
\end{equation}
with $\vb*{\delta}_j$ a lattice vector and the sum over the $\mu$\textsuperscript{th} nearest neighbors. The spin-wave spectrum is then determined from the eigenvalues of the matrix in Eq. \eqref{eq:heisenbergEq},
\begin{equation}
    \omega^2(\vb{k}) = 16 S^2 [\Tilde{J}^2(\vb{k}) - J_1^2 (\vb{k}) ] .
\end{equation}
Expanding the dispersion as a power series in $\vb{k}$ then yields the mapping between the lattice and the low energy/long wavelength effective field theory parameters
\begin{subequations}
\begin{align}
    \chi_\perp &= \frac{1}{8 J_1 a^2} ,\\
    \frac{\rho}{J_1 S^2} &=  1 - \frac{2 J_2}{J_1} - \frac{4 J_3}{J_1} ,\\
    \frac{b_1}{J_1 S^2 a^2} &=  -\frac{5}{24} +  \frac{J_2}{6 J_1} + \frac{4 J_3}{3 J_1} + 2 \left( \frac{J_2}{2 J_1} + \frac{J_3}{J_1} \right)^2  , \\
    \frac{b_2}{J_1 S^2 a^2} &= - \frac{1}{8} +  \frac{J_2}{2 J_1} + 2 \left( \frac{J_2}{2 J_1} + \frac{J_3}{J_1} \right)^2 .
\end{align}
\end{subequations}

\section{Momentum Shell Renormalization Group\label{app:RG}}

In this appendix, we derive the RG beta function given in Eq. \eqref{eq:kshellbetafunction} using the momentum shell method, and provide a more accurate estimate of the size and location of the gapped phase. 

We start from the low energy effective theory given by the Lagrangian
\begin{align}
    \mathcal{L} = \frac{\chi_\perp}{2} (\del_t \Vec{n})^2 - \frac{\rho}{2} (\grad \Vec{n})^2 - \frac{b_1}{2}  \left[(\partial_x^2 \Vec{n})^2 + (\partial_y^2 \Vec{n})^2\right] \\
    - b_2(\partial_x^2 \Vec{n})\cdot(\partial_y^2 \Vec{n}), \nonumber
\end{align}
and define the dimensionless couplings $u = \Lambda^{2-z} \chi_\perp^{-1}$, $\varrho = \Lambda^{2-2z} \rho \chi_\perp^{-1}$, $\beta_{1,2} = \Lambda^{4-2z} b_{1,2} \chi_\perp^{-1}$, where $z$ is an unspecified dynamical exponent denoting the relative scaling of space and time dimensions and $\Lambda \sim \pi/a$ is the ultraviolet momentum cutoff scale; note that in the renormalization scheme we use here there is no frequency cutoff. Next, we define $\Vec{n} = (\Vec{\pi},\sigma)$ and expand around the N\'eel state where $\langle \Vec{n}\rangle = (0,\langle \sigma \rangle)$ and the variance of the transverse Goldstone modes is assumed to be small. Then, integrating out the momenta in a shell $s \Lambda < \abs{\vb{k}} < \Lambda$ yields the RG transformations for the dimensionless couplings
\begin{subequations}
\begin{align}
    \frac{1}{u'} = \frac{s^{z-d}}{u}  \left( 1 - \frac{N-2}{N-1} \langle \vec{\pi}^2 \rangle \right), \\
    \varrho' = s^{2-2z}\left( \varrho + \Lambda^{-2} \frac{3 \beta_1 + \beta_2}{N-1} \langle (\grad \vec{\pi})^2 \rangle \right), \\
    \beta_{1,2}' = s^{4-2z} \beta_{1,2} .
\end{align}
\end{subequations}
As explained in the main text, in the vicinity of the LP where $\varrho \ll \beta_{1,2}$, the Goldstone modes within the momentum shell have a dispersion scaling as $\omega_{\vb{q}} \simeq \sqrt{b_{1,2}/\chi_\perp} \vb{q}^2$ and so the corresponding dynamical exponent is $z=2$. It follows that $u$ and $\beta_{1,2}$ are marginal operators (the former being marginally relevant) while $\varrho$ is an infrared relevant perturbation. In this regime, it is straightforward to perform the momentum shell integral to find
\begin{align}
    \langle \vec{\pi}^2 \rangle &\simeq (N-1) \frac{u}{2\pi^2 \sqrt{\beta_1}} K(\Delta) \delta\ell,
\end{align}
where we have defined $s = 1 - \delta\ell$, assuming that $\delta \ell \ll 1$, $K(z)$ is the complete elliptic integral of the first kind, and $\Delta = (\beta_1 - \beta_2)/2\beta_1$. The analog of the result given in Ref. \cite{Ioffe1988} follows from ignoring the flow of $\varrho$, in which case to leading order in the couplings, we obtain the one-loop beta function for $u$
\begin{equation}
    \frac{d u}{d \ell} =   \frac{(N-2) K(\Delta) u^2}{2\pi^2 \sqrt{\beta_1}},
\end{equation}
as given in the main text. 

However, the simple analysis above is complicated by the non-trivial flow of the dimensionless stiffness $\varrho$. To the best of our knowledge, this has not been discussed in previous literature. If we wish to work to quadratic order in the couplings, it is necessary to expand $\langle (\grad \vec{\pi})^2 \rangle $ to first order in $\varrho$,
\begin{equation}
    \langle (\grad \vec{\pi})^2 \rangle \simeq (N-1) \Lambda^{2} \frac{u}{2\pi^2 \sqrt{\beta_1}} \left[ K(\Delta) -  \frac{E(\Delta) \varrho}{\beta_1 + \beta_2} \right]\delta\ell,
\end{equation}
where $E(z)$ is the complete elliptic integral of the second kind. This yields the one-loop beta function for $\varrho$
\begin{equation}
    \frac{d \varrho}{d\ell} = 2\varrho + \left(3\beta_1 + \beta_2\right) \frac{u}{2\pi^2 \sqrt{\beta_1}} \left[ K(\Delta) -  \frac{E(\Delta) \varrho}{\beta_1 + \beta_2} \right] .
\end{equation}
Since $u$ is marginal and $\varrho$ is relevant, it is clear that, for generic (but small) values of $u(0)$ and $\varrho(0)$, the former flows much more slowly under the RG than the latter. Therefore, assuming that $u(\ell)$ is approximately constant over the flow of $\varrho(\ell)$, we can integrate this beta function to find that the flow is influenced by an unstable quasi-fixed trajectory $\varrho_*(u)$,
\begin{align}
    \varrho(\ell) - \varrho_*(u) &\simeq [\varrho(0) - \varrho_*(u)] \\
    &\times\exp\left( \left[2 - \frac{(3\beta_1 + \beta_2) E(\Delta)}{2\pi^2(\beta_1 + \beta_2)} u(\ell) \right] \ell \right), \nonumber
\end{align}
where
\begin{align}
    \varrho_*(u) &= \frac{K(\Delta)u}{[E(\Delta) u - 4\pi^2 \sqrt{\beta_1}/(3\beta_1 + \beta_2)]} \nn \\
    &\simeq - \frac{(3\beta_1 + \beta_2) K(\Delta) u}{4\pi^2 \sqrt{\beta_1}},
\end{align}
to leading order in $u$. From this we see that proximity of $\rho(0)$ to $\rho_*(u)$ will drastically slow down the flow of the stiffness. Then, we expect that for $\abs{\varrho(0) - \varrho_*(u)}$ large, $\abs{\varrho(\ell)}$ will become $\mathcal{O}(1)$ before $u(\ell)$, in which case the expansion around the LP breaks down and the system crosses over into the $z=1$ dynamical exponent regime of the conventional NLSM. On the other hand, for $\abs{\varrho(0) - \varrho_*(u)}$ small enough, $u(\ell)$ will become $\mathcal{O}(1)$ before $\abs{\varrho(\ell)}$, in which case we expect the system to form a gap due to dynamical mass generation. By integrating the RG flow equations exactly, we find that the tolerance for deviation of the bare stiffness from the fixed trajectory must be exponentially small in order to be in the magnetically gapped phase formed due to strong $z=2$ quantum fluctuations,
\begin{equation}
    \abs{\varrho_c(0) - \varrho_*(u(0))} \propto \exp\left(- \frac{4\pi^2 \sqrt{\beta_1}}{(N-2) K(\Delta) u(0)} \right),
\end{equation}
where $\rho_c(0)$ is the critical value of the bare stiffness and the qualitatively unimportant prefactor of the exponential is determined by sub-leading logarithmic corrections. This result agrees with the expectations from our more qualitative estimate discussed in the main text, though now the exponent agrees exactly with the large-$N$ result presented in Sec. \ref{sec:largeNv1}. Admittedly, the fixed trajectory for $\varrho(\ell)$ is strictly negative $\varrho_*(u) < 0$ for all physical values of $u > 0$. Therefore, one may prefer to repeat this RG analysis by instead considering a manifold of Goldstone modes corresponding to a spiral phase. However, since we are working sufficiently close to the LP such that $\abs{\varrho} \ll \beta_{1,2}$, the conclusions are fundamentally unchanged.

\begin{widetext}

\section{Saddle-Point Equations \label{app:saddles}}

First, we consider the effective potential for the classical $(2+0)$-dimensional field theory
\begin{equation}
    \Gamma[\Psi_1,\Psi_2] = \frac{\gamma_1 \Psi_1^2}{2} + \frac{\gamma_2 \Psi_2^2}{2}  + \Psi_1 \psi_1 + \Psi_2 \psi_2   - \frac{\lambda}{2} - \frac{(\eta - \rho)^2}{2\gamma_0} +\frac{T}{4} \int \frac{d^2 \vb{q}}{(2\pi)^2} \ln \left( [\eta \vb{q}^2 + b_{ij} q_i^2 q_j^2 + \lambda]^2 - [\psi_{ij}q_iq_j]^2 \right).
\end{equation}
The momentum integral is ultraviolet divergent and hence, the theory requires renormalization to obtain the physical potential. First, the naive ultraviolet divergence should be subtracted,
\begin{equation}
    \Gamma[\Psi_1,\Psi_2] \to \Gamma[\Psi_1,\Psi_2] - \frac{T}{4} \int \frac{d^2 \vb{q}}{(2\pi)^2} \ln ([b_{ij} q_i^2 q_j^2]^2),
\end{equation}
since it does not depend on any of the variational parameters. However, a divergence remains which must be cured by renormalizing the stiffness parameter:
\begin{equation}
    \rho_R = \rho + \frac{T \gamma_0}{2} \int \frac{d^2 \vb{q}}{(2\pi)^2} \frac{\vb{q}^2}{b_{ij} q_i^2 q_j^2 + \mu^2},
\end{equation}
where $\mu$ is the renormalization scale. Using these definitions, the momentum integral can be evaluated to obtain the potential
\begin{align}
    \Gamma[\Psi_1,\Psi_2] = \frac{\gamma_1 \Psi_1^2}{2} + \frac{\gamma_2 \Psi_2^2}{2}  + \Psi_1 \psi_1 + \Psi_2 \psi_2   - \frac{\lambda}{2} - \frac{(\eta - \rho_R)^2}{2\gamma_0} + T \int_0^{2\pi} \frac{d \theta}{2\pi} \frac{\eta}{16\pi b_\theta} \left[ 2 - \ln\left(\frac{\lambda}{\mu^2}\right) \right] \\
    + T \int_0^{2\pi} \frac{d \theta}{2\pi} \frac{1}{16\pi b_\theta} \left[ \sqrt{4 b_\theta  \lambda-(\eta +\psi_\theta)^2} \cos ^{-1}\left(\frac{\eta +\psi_\theta}{\sqrt{4 b_\theta \lambda}}\right) + \sqrt{4 b_\theta  \lambda-(\eta -\psi_\theta)^2} \cos ^{-1}\left(\frac{\eta -\psi_\theta}{\sqrt{4 b_\theta \lambda}}\right)  \right] \nonumber
\end{align}
where we have defined $b_\theta = b_1(\cos^4\theta + \sin^4\theta) + 2b_2 \cos^2\theta\sin^2\theta$ and $\psi_\theta = \psi_1 \cos(2\theta) + \psi_2 \sin(2\theta)$; the remaining angular integral cannot be performed analytically. It follows that the renormalized saddle-point equations for the variational parameters $\lambda$, $\eta$, $\psi_1$ and $\psi_2$ are, respectively,
\begin{subequations}
\begin{align}
    &\int_0^{2\pi} \frac{d\theta}{2\pi} \left[ \frac{\cos ^{-1}\left([\eta +\psi_\theta]/\sqrt{4 b_\theta \lambda}\right)}{4\pi\sqrt{4 b_\theta  \lambda-(\eta +\psi_\theta)^2}}   + \frac{\cos ^{-1}\left([\eta - \psi_\theta]/\sqrt{4 b_\theta \lambda}\right)}{4\pi\sqrt{4 b_\theta  \lambda-(\eta -\psi_\theta)^2}}  \right] = \frac{1}{T}, \\
    &\int_0^{2\pi} \frac{d \theta}{2\pi} \frac{1}{4  b_\theta} \left[   \frac{(\eta + \psi_\theta) \cos ^{-1}\left([\eta +\psi_\theta]/\sqrt{4 b_\theta \lambda}\right)}{4\pi\sqrt{4 b_\theta  \lambda-(\eta +\psi_\theta)^2}}   +  \frac{(\eta-\psi_\theta) \cos ^{-1}\left([\eta - \psi_\theta]/\sqrt{4 b_\theta \lambda}\right)}{4\pi\sqrt{4 b_\theta  \lambda-(\eta -\psi_\theta)^2}} + \ln \left(\frac{\lambda}{\mu^2}\right) \right] = \frac{\rho_R - \eta}{T \gamma_0},  \\
    &\int_0^{2\pi} \frac{d \theta}{2\pi} \frac{\cos(2\theta)}{4  b_\theta} \left[  \frac{(\eta + \psi_\theta) \cos ^{-1}\left([\eta +\psi_\theta]/\sqrt{4 b_\theta \lambda}\right)}{4\pi\sqrt{4 b_\theta  \lambda-(\eta +\psi_\theta)^2}}   - \frac{(\eta-\psi_\theta)\cos ^{-1}\left([\eta - \psi_\theta]/\sqrt{4 b_\theta \lambda}\right)}{4\pi\sqrt{4 b_\theta  \lambda-(\eta -\psi_\theta)^2}}  \right] = \frac{\Psi_1}{T},  \\
    &\int_0^{2\pi} \frac{d \theta}{2\pi} \frac{\sin(2\theta)}{4  b_\theta} \left[  \frac{(\eta + \psi_\theta) \cos ^{-1}\left([\eta +\psi_\theta]/\sqrt{4 b_\theta \lambda}\right)}{4\pi\sqrt{4 b_\theta  \lambda-(\eta +\psi_\theta)^2}}   -  \frac{(\eta-\psi_\theta) \cos ^{-1}\left([\eta - \psi_\theta]/\sqrt{4 b_\theta \lambda}\right)}{4\pi\sqrt{4 b_\theta  \lambda-(\eta -\psi_\theta)^2}}  \right] = \frac{\Psi_2}{T} .
\end{align}
\end{subequations}

Next, we consider the effective potential for the quantum $(2+1)$-dimensional field theory at zero temperature
\begin{align}
    \Gamma[\Psi_1,\Psi_2] &= \frac{\gamma_1 \Psi_1^2}{2} + \frac{\gamma_2 \Psi_2^2}{2}  + \Psi_1 \psi_1 + \Psi_2 \psi_2 - \frac{(\eta - \rho)^2}{2\gamma_0} - \frac{\lambda}{2} + \frac{1}{2} \sum_{\ell=1}^2 \sigma_\ell \left( -\eta \nabla^2 - \psi_{ij} \del_i \del_j + b_{ij} \del_i^2 \del_j^2 + \lambda \right) \sigma_\ell \\
    &+\frac{g}{4} \int \frac{d^2 \vb{q}}{(2\pi)^2} \frac{d \omega}{2\pi} \ln \left( [\omega^2 + \eta \vb{q}^2 + b_{ij} q_i^2 q_j^2 +  \lambda]^2 - [\psi_{ij}q_i q_j]^2 \right)  . \nonumber
\end{align}
The frequency integral can be performed analytically after subtracting a trivial ultraviolet divergence
\begin{equation}
    \Gamma[\Psi_1,\Psi_2] \to \Gamma[\Psi_1,\Psi_2] - \frac{g}{4} \int \frac{d^2 \vb{q}}{(2\pi)^2} \frac{d \omega}{2\pi} \ln([\omega^2 + b_{ij} q_i^2 q_j^2]^2),
\end{equation}
in which case we obtain
\begin{align}
    \Gamma[\Psi_1,\Psi_2] &= \frac{\gamma_1 \Psi_1^2}{2} + \frac{\gamma_2 \Psi_2^2}{2}  + \Psi_1 \psi_1 + \Psi_2 \psi_2 - \frac{(\eta - \rho)^2}{2\gamma_0} - \frac{\lambda}{2} + \frac{1}{2} \sum_{\ell=1}^2 \sigma_\ell \left( -\eta \nabla^2 - \psi_{ij} \del_i \del_j + b_{ij} \del_i^2 \del_j^2 + \lambda \right) \sigma_\ell \\
    &+\frac{g}{4} \int \frac{d^2 \vb{q}}{(2\pi)^2} \frac{d \omega}{2\pi} \left( \sqrt{\eta \vb{q}^2 + \psi_{ij}q_i q_j + b_{ij} q_i^2 q_j^2 +  \lambda} + \sqrt{\eta \vb{q}^2 - \psi_{ij}q_i q_j + b_{ij} q_i^2 q_j^2 +  \lambda}  - 2 \sqrt{b_{ij}q_i^2 q_j^2} \right)  . \nonumber
\end{align}
As above, there are remaining ultraviolet divergences which must be cured by renormalizing $g$, $\rho$, $\gamma_0$, $\Psi_1$ and $\Psi_2$, respectively,
\begin{subequations}
\begin{align}
    \frac{1}{g_R} &= \frac{1}{g} - \int \frac{d^2 \vb{q}}{(2\pi)^2} \frac{1}{2 \sqrt{\smash[b]{b_{ij}q_i^2 q_j^2 + \mu^2}}} ,\\
    \frac{\rho_R}{g_R \gamma_{0,R}} &= \frac{\rho}{g \gamma_0} + \frac{1}{2} \int \frac{d^2 \vb{q}}{(2\pi)^2} \frac{\vb{q}^2}{2 \sqrt{\smash[b]{b_{ij}q_i^2 q_j^2 + \mu^2}}}, \\
    \frac{1}{g_R \gamma_{0,R}} &= \frac{1}{g \gamma_0} + \frac{1}{2} \int \frac{d^2 \vb{q}}{(2\pi)^2}  \frac{(\vb{q}^2)^2}{4( b_{ij}q_i^2 q_j^2+ \mu^2)^{3/2}}, \\
    \frac{\Psi_{1,R}}{g_R} &= \frac{\Psi_1}{g} + \frac{\psi_1}{2} \int \frac{d^2 \vb{q}}{(2\pi)^2}  \frac{(q_x^2 - q_y^2)^2}{4( b_{ij}q_i^2 q_j^2+ \mu^2)^{3/2}}, \\
    \frac{\Psi_{2,R}}{g_R} &= \frac{\Psi_2}{g} + \frac{\psi_2}{2} \int \frac{d^2 \vb{q}}{(2\pi)^2}  \frac{(2 q_x q_y)^2}{4( b_{ij}q_i^2 q_j^2+ \mu^2)^{3/2}},
\end{align}
\end{subequations}
and $\gamma_{1,R}\Psi_{1,R}^2/g_R = \gamma_1 \Psi_1^2/g$, $\gamma_{2,R}\Psi_{2,R}^2/g_R = \gamma_2 \Psi_2^2/g$, and $\sigma_{i,R}^2/g_R = \sigma_{i}^2/g$, where $\mu$ is the renormalization scale. The physical effective potential can then be evaluated (suppressing $R$ subscripts for clarity),
\begin{align}
    &\Gamma[\Psi_1,\Psi_2] = \frac{\gamma_{1} \Psi_1^2}{2} + \frac{\gamma_2 \Psi_2^2}{2}  + \Psi_1 \psi_1 + \Psi_2 \psi_2   - \frac{\lambda}{2} - \frac{(\eta - \rho)^2}{2\gamma_0} \\
    &+ \frac{1}{2} \sum_{\ell=1}^2 \sigma_\ell \left( -\eta \nabla^2 - \psi_{ij} \del_i \del_j + b_{ij} \del_i^2 \del_j^2 + \lambda \right) \sigma_\ell + g \int_0^{2\pi} \frac{d\theta}{2\pi} \frac{1}{128 \pi b_\theta^{3/2}} \left[ 4 b_\theta \lambda - \eta^2 - \psi_\theta^2 - 4\eta\sqrt{b_\theta}(\sqrt{\lambda} - 2\mu)     \right] \nonumber \\
    &+ g \int_0^{2\pi} \frac{d\theta}{2\pi} \frac{1}{128 \pi b_\theta^{3/2}} \left[ (4b_\theta \lambda - (\eta + \psi_\theta)^2) \ln\left(\frac{\sqrt{4b_\theta\mu^2}}{\sqrt{4b_\theta\lambda} + (\eta + \psi_\theta)} \right) + (4b_\theta \lambda - (\eta - \psi_\theta)^2) \ln\left(\frac{\sqrt{4b_\theta\mu^2}}{\sqrt{4b_\theta\lambda} + (\eta - \psi_\theta)} \right) \right] \nonumber
\end{align}
It follows that the renormalized saddle-point equations are
\begin{subequations}
\begin{align}
    &\int_0^{2\pi} \frac{d\theta}{2\pi} \frac{1}{16\pi \sqrt{b_\theta}}  \left[ \ln\left(\frac{\sqrt{4b_\theta\mu^2}}{\sqrt{4b_\theta\lambda} + (\eta + \psi_\theta)} \right) + \ln\left(\frac{\sqrt{4b_\theta\mu^2}}{\sqrt{4b_\theta\lambda} + (\eta - \psi_\theta)} \right)  \right]  = \frac{1}{g} - \frac{\sigma_{i}^2}{g}, \\
    &\int_0^{2\pi} \frac{d \theta}{2\pi} \frac{1}{64 \pi b_\theta^{3/2}} \left[ (\eta + \psi_\theta) \ln\left(\frac{\sqrt{4b_\theta\mu^2}}{\sqrt{4b_\theta\lambda} + (\eta + \psi_\theta)} \right) + (\eta - \psi_\theta) \ln\left(\frac{\sqrt{4b_\theta\mu^2}}{\sqrt{4b_\theta\lambda} + (\eta - \psi_\theta)} \right)  + 4 \sqrt{b_\theta} (\sqrt{\lambda} - \mu) \right]  \\
    &\kern2em = \frac{(\rho - \eta)}{g \gamma_{0}} - \frac{1}{2 g} \sigma_{i} \nabla^2 \sigma_{i}, \nonumber \\
    &\int_0^{2\pi} \frac{d \theta}{2\pi} \frac{\cos(2\theta)}{64 \pi b_\theta^{3/2}} \left[ (\eta + \psi_\theta) \ln\left(\frac{\sqrt{4b_\theta\mu^2}}{\sqrt{4b_\theta\lambda} + (\eta + \psi_\theta)} \right) - (\eta - \psi_\theta) \ln\left(\frac{\sqrt{4b_\theta\mu^2}}{\sqrt{4b_\theta\lambda} + (\eta - \psi_\theta)} \right)  \right] \\
    &\kern2em = \frac{\Psi_{1}}{g} - \frac{1}{2 g}  \sigma_{i} (\del_x^2 - \del_y^2) \sigma_{i}, \nonumber \\
    &\int_0^{2\pi}  \frac{d \theta}{2\pi} \frac{\sin(2\theta)}{64 \pi b_\theta^{3/2}} \left[ (\eta + \psi_\theta) \ln\left(\frac{\sqrt{4b_\theta\mu^2}}{\sqrt{4b_\theta\lambda} + (\eta + \psi_\theta)} \right) - (\eta - \psi_\theta) \ln\left(\frac{\sqrt{4b_\theta\mu^2}}{\sqrt{4b_\theta\lambda} + (\eta - \psi_\theta)} \right)  \right] \\
    &\kern2em = \frac{\Psi_{2}}{g} - \frac{1}{g}  \sigma_{i} \del_x \del_y \sigma_{i}, \nonumber \\
    &(-\eta \nabla^2 - \psi_{ij} \del_i \del_j + b_{ij} \del_i^2 \del_j^2 + \lambda) \sigma_{\ell} = 0 ,
\end{align}
\end{subequations}
where summation over all indices $i$, $j$ is implied, and as before, the remaining angular integrals cannot be performed analytically.

\end{widetext}

\bibliography{refs}

\end{document}